\documentclass[twocolumn]{aastex63}

\hypersetup{linkcolor=red,citecolor=blue,filecolor=cyan,urlcolor=blue}
\usepackage{graphicx,color}
\usepackage{amssymb}
\usepackage{amsmath}

\usepackage{natbib}
\usepackage{txfonts}
\usepackage{multirow}
\usepackage{array}
\usepackage{rotating}
\usepackage{sidecap}
\usepackage{hyperref}
\usepackage{epstopdf}
\usepackage{footnote}
\usepackage{tabularx}
\usepackage{booktabs}
\usepackage{longtable}
\usepackage{tabu}
\usepackage{longtable}

\newcommand{\kms}{km~s$^{-1}$}
\newcommand{\degree}{\ensuremath{^\circ}}

\newcommand{\YOHKOH}{\textit{Yohkoh}}
\newcommand{\Hinode}{\textit{Hinode}}

\newcommand{\sdo}{\textit{SDO}}

\shortauthors{Panesar et al.}

\shorttitle{Jet Magnetic Explosion Onset}
\shortauthors{Panesar et al.}

\begin{document}
	
	\title{\bf Onset of Magnetic Explosion in Solar Coronal Jets in Quiet Regions on the Central Disk}


	\author[0000-0001-7620-362X]{Navdeep K. Panesar}
	
	\affil{Lockheed Martin Solar and Astrophysics Laboratory, 3251 Hanover Street, Bldg. 252, Palo Alto, CA 94304, USA}
	\affil{Bay Area Environmental Research Institute, NASA Research Park, Moffett Field, CA 94035, USA}
	\email{panesar@lmsal.com; ronald.l.moore@nasa.gov}

	\author[0000-0002-5691-6152]{Ronald L. Moore}
	\affiliation{NASA Marshall Space Flight Center, Huntsville, AL 35812, USA}
	\affiliation{Center for Space Plasma and Aeronomic Research (CSPAR), UAH, Huntsville, AL 35805, USA}
	
	\author[0000-0003-1281-897X]{Alphonse C. Sterling}	
	\affiliation{NASA Marshall Space Flight Center, Huntsville, AL 35812, USA}

	\begin{abstract}
		
		     We examine the initiation of 10 coronal jet eruptions in quiet regions on the central disk, thereby avoiding near-limb spicule-forest obscuration of the slow-rise onset of the minifilament eruption.  From the  \textit{\sdo}/AIA 171 \AA\ 12-second-cadence movie of each eruption, we (1) find and compare the start times of the minifilament’s slow rise, the jet-base bright point, the jet-base interior brightening, and the jet spire, and (2) measure the minifilament’s speed at the start and end of its slow rise.  From (a) these data, (b) prior observations showing that each eruption was triggered by magnetic flux cancelation under the minifilament, and (c) the breakout-reconnection current sheet observed in one eruption, we confirm that quiet-region jet-making minifilament eruptions are miniature versions of CME-making filament eruptions, and surmise that in most quiet-region jets: (1)  the eruption starts before runaway reconnection starts, (2) runaway reconnection does not start until the slow-rise speed is at least $\sim$ 1 \kms, and (3) at and before eruption onset there is no current sheet of appreciable extent.  We therefore expect: (i) many CME-making filament eruptions are triggered by flux cancelation under the filament, (ii) emerging bipoles seldom, if ever, directly drive jet production because the emergence is seldom, if ever, fast enough, and (iii) at a separatrix or quasi-separatrix in any astrophysical setting of magnetic field in low-beta plasma, a current sheet of appreciable extent can be built only dynamically by a magnetohydrodynamic convulsion of the field, not by quasi-static gradual converging of the field.

	\end{abstract}

	\section{INTRODUCTION}
	
This paper investigates the initiation of typical solar coronal jets that occur in quiet regions and coronal holes.  Such events are reviewed by \cite{shibata11} and \cite{raouafi16}.
		
		Coronal jets were discovered in coronal X-ray images from \YOHKOH\ \citep{shibata92}.   They are elongated features that erupt into  the solar corona \citep{innes16,raouafi16}. They occur all over the Sun: in coronal holes, quiet regions, and in and around the edges of the active regions \citep{shibata92,cirtain07,savcheva07,nistico09,schmieder13,sterling16,panesar16a,chandra17}. They have lengths ranging from less than 10$^4$ km to more than  10$^5$ km and have been reported to have lifetimes ranging from minutes to hours  \citep{shimojo96,panesar18b,panesar19,tiwari19}.  A typical coronal jet is  a brief ($\sim$ 10 min) spire that extends from a broader base.  The spire is usually accompanied in its base by a transient bright feature called the jet-base bright point or jet bright point.  The jet bright point is inside an edge of the base, much smaller than the base, and brighter than most of the rest of the base \citep{shibata92}. The base is often visible in coronal X-ray images as a so-called X-ray bright point for hours before and after the jet event, often appears to be a compact cluster of bright loops, and is usually less than 30,000 km in diameter.  The term X-ray bright point refers to the whole long-lasting bright cluster, not to the much smaller jet bright point that occurs at an edge of the cluster during a jet.
		
		Before \YOHKOH, registration of photospheric magnetograms with coronal X-ray images from Skylab had established that X-ray bright points are the coronal X-ray signatures of so-called ephemeral regions \citep{golub77}.  An ephemeral region is a miniature bipolar solar active region, closed-loop magnetic field that emerges into the solar atmosphere in a fraction of a day and decays to extinction in $\sim$ 1 day \citep[e.g.][]{bruzek77}.  If the ephemeral-region bipole emerges in an area of predominantly unipolar magnetic flux, i.e., in the feet of unipolar open or far-reaching magnetic field, the closed field in the resulting X-ray bright point is expected to have anemone magnetic form.  The expected anemone field consists of two sets of magnetic loops: (1) emerged loops of the emerging bipole, and (2) loops made by reconnection of emerged loops with ambient oppositely-directed far-reaching field, loops connecting ambient flux to emerged flux of the opposite polarity.  The \YOHKOH\ coronal X-ray images had enough resolution to show that X-ray bright points commonly do have anemone form \citep[e.g.][]{acton92}.


\begin{deluxetable*}{c c c c c c c c c c}
	\tablewidth{0pt}
	\tabletypesize{\footnotesize}
	\renewcommand{\arraystretch}{1.0}
	\tablenum{1}
	\tablecaption{Key Data for our 10 Central-Disk Quiet-Region Jet Eruption Onsets \label{tab:list}}
	\centerwidetable
	\tablehead{
		\colhead{Jet} & 
		\colhead{Date} & 
		\colhead{Location\tablenotemark{\scriptsize a}} & 
		\colhead{MF slow-rise\tablenotemark{\scriptsize b}} & 
		\colhead{MF slow-rise\tablenotemark{\scriptsize c}} & 
		\colhead{MF slow-rise\tablenotemark{\scriptsize d}} &
		\colhead{MF slow-rise\tablenotemark{\scriptsize e}} & 
		\colhead{JBP start\tablenotemark{\scriptsize f}} & 
		\colhead{BIB start\tablenotemark{\scriptsize g}} &
		\colhead{Spire start\tablenotemark{\scriptsize h}}
		\\\colhead{No.}&
		\colhead{} & 
		\colhead{x, y (arcsec)}&
		\colhead{start time (UT)}&
		\colhead{end time (UT)}&
		\colhead{start speed (\kms)}&
		\colhead{end speed (\kms)}&
		\colhead{time (UT)}&
		\colhead{time (UT)}&
		\colhead{time (UT)}}
	\startdata
	J1   & 2012 Mar 22 & --470, --100  & 04:41:00 $\pm$60s  & 04:46:00$\pm$12s   &   1.2$\pm$0.3   & 6.0$\pm$1.0   &   04:43:24 $\pm$12s  & 04:48:00 $\pm$24s & 04:51:48 $+$60s \\   
	J2   &  2012 Jul 01  &  --44, 285  & 08:19:11 $\pm$60s  & 08:27:00$\pm$30s  &  1.4$\pm$0.5  &  5.0$\pm$2.0  &  08:21:35 $\pm$60s   &  08:19:23$\pm$12s     &  08:31:59 $\pm$12s    \\   
	J3   & 2012 Jul 07 & --192, --180 & 21:13:00$ \pm$60s  & 21:33:00$\pm$60s  &  0.5$\pm$0.3  & 4.0$\pm$2.0   &  21:17:35 $\pm$12s   &  21:32:59 $\pm$24s &  21:32:59 $\pm$24s  \\   
	J4   & 2012 Aug 05 & --485, 190  & 01:35:00$\pm$60s   & 01:47:00$\pm$60s  &  7.5$\pm$0.5  & 18.5$\pm$1.0  &  01:39:47 $\pm$12s & 01:43:37 $\pm$12s  &  01:41:41 $\pm$24s  \\   
	J5   & 2012 Aug 10 & --168, --443  & 22:50:00 $\pm$60s  & 23:02:00$\pm$60s  &  1.5$\pm$0.5  & 9.0$\pm$1.0   &  22:59:59 $\pm$12s &22:53:47 $\pm$12s   &  23:05:59 $\pm$72s  \\ 
	J6   & 2012 Sept 20  & --158, --486  & 22:34:00 $\pm$60s  & 22:58:00$\pm$60s  &  1.2$\pm$0.5   & 4.5$\pm$1.0  &  22:46:47 $\pm$12s&22:49:59 $\pm$12s   &  22:56:47 $\pm$12s    \\  
	J7   & 2012 Sept 21   & --115, --485  & 03:27:11 $\pm$90s   & 03:31:00$\pm$30s  &  3.0$\pm$0.5  & 17.0$\pm$1.0   &  03:28:11 $\pm$24s   & 03:31:11 $\pm$12s  & 03:33:11 $\pm$24s  \\   
	J8   & 2012 Sept 22   &  --338, 103  & 01:13:00 $\pm$60s  & 01:28:00$\pm$30s   &  1.0$\pm$ 0.3  & 14.5$\pm$1.0   &  01:24:35 $\pm$12s & 01:19:35 $\pm$24s  & 01:26:23 $\pm$12s  \\ 
	J9   & 2012 Nov 13   & --28, --307  & 03:50:35 $\pm$24s   & 04:20:00$\pm$60s &  1.2$\pm$0.5  & 4.5$\pm$0.5  &  04:20:11 $\pm$24s   &  03:50:35 $\pm$24s  &   04:14:36 $\pm$12s   \\ 
	J10  & 2012 Dec 13   & 26, 50  &   10:27:11 $\pm$24s   & 10:35:30$\pm$30s &  3.5$\pm$1.0  & 8.0$\pm$1.0    &  10:29:59 $\pm$48s   &  10:34:23 $\pm$60s  &   10:35:47 $\pm$24s   \\ 
	\noalign{\smallskip}\tableline\tableline \noalign{\smallskip}
	average$\pm$1$\sigma$$_{ave}$ &  &  &  &  & 2.2$\pm$2.0  & 9.0$\pm$5.5 & &  & \\
	\hline
	\enddata
	\singlespace
	\tablecomments{
		\\\textsuperscript{a} Approximate location of the jet’s base on the solar disk.
		\\\textsuperscript{b}Time of first detection of the slow rise of the minifilament (MF).
		\\\textsuperscript{c}Time of the end of the slow rise of the minifilament (MF) (= time of the start of the fast rise of the minifilament).
		\\\textsuperscript{d}Minifilament’s proper-motion speed at the start of the minifilament’s slow rise.
		\\\textsuperscript{e}Minifilament’s proper-motion rise speed at the end of the minifilament’s slow rise (at the start of the minifilaments’s fast rise).
		\\\textsuperscript{f}Time of the first detection of the jet bright point (JBP) under the minifilament.
		\\\textsuperscript{g}Time of the first detection of the jet’s base interior brightening; BIB  (BIB; also known as external brightening, \citealt{panesar16b}).
		\\\textsuperscript{h}Time of the first detection of the jet spire.
	}
	
\end{deluxetable*}
		
	  From (1) the Skylab finding that X-ray bright points mark ephemeral regions, (2) the \YOHKOH\ observation that the closed magnetic field constituting an X-ray bright point often has anemone form, and (3) the \YOHKOH\ observation that X-ray jet spires shoot out from X-ray bright points, it is plausible that (1) the jet’s spire and jet bright point are produced together by a burst of reconnection of the ephemeral region’s emerging closed field with ambient oppositely-directed far-reaching field\footnote{The spire would be injected reconnection-heated plasma on the reconnected far-reaching field.  The jet bright point would be reconnection-heated plasma in the corresponding closed loops made by the reconnection.  That is, the jet bright point would be hot new loops added to the set of anemone loops connecting nearby ambient flux to the ephemeral region’s opposite-polarity flux.}, and (2) the burst of reconnection results from the emerging magnetic arch pushing against the far-reaching field, building between them the current sheet at which the reconnection occurs.  This concept for X-ray jet production was modeled by \cite{yokoyama95} with 2D MHD simulations of current-sheet formation and reconnection driven by a buoyant magnetic arch as it wells up through the photosphere into the low corona through ambient unipolar far-reaching field.  These simulations demonstrated the physical feasibility of the concept and produced jet spires and jet bright points in reasonable agreement with \YOHKOH\ coronal X-ray images of jets.

		     The \Hinode\ X-Ray Telescope (XRT) has provided coronal X-ray images having higher resolution and faster cadence than those from \YOHKOH.  From XRT images together with EUV images from the \textit{Solar Dynamics Observatory} (\sdo; \citealt{pensell12})/Atmospheric Imaging Assembly (AIA), \cite{sterling15} found that coronal X-ray jets in polar regions are usually not directly made by an emerging bipole in the manner simulated by \cite{yokoyama95}.  Instead, most are made by a minifilament eruption that appears to be a miniature version of either (1) a blowout (ejective) filament eruption that makes a flare arcade in tandem with a coronal mass ejection (CME) or (2) a confined filament eruption that makes a flare without a CME.  In the same way as a filament in the core of a larger magnetic arcade \citep{Martin73,Martin98}, the pre-eruption minifilament runs along the polarity inversion line (PIL) in the core of a lobe of the jet-base magnetic anemone and hence is a sign that the core field of that lobe is greatly sheared and might be partly a flux rope.  All of the magnetic field in that lobe participates in the minifilament eruption: the lobe erupts outward as a whole with the erupting minifilament flux rope inside it.  If the eruption is a blowout eruption, the lobe and minifilament erupt well up into the ambient far-reaching field.  If the eruption is a confined eruption, the eruption stops when the lobe has distended only a fraction of its initial extent, and the erupting minifilament is largely confined within the jet-base anemone.  In either case, because the jet bright point sits on the PIL under the erupting minifilament, the jet bright point is evidently made by internal reconnection of the erupting lobe’s opposite-polarity legs that implode together under the erupting minifilament flux rope as in larger filament eruptions \citep[e.g.][]{moore01}, not by reconnection of closed field with far-reaching field as in the jet model of \cite{yokoyama95}.  The jet spire is evidently made by external reconnection of the erupting lobe with encountered far-reaching ambient field that is rooted just outside of the rest of the jet-base anemone, that is, just outside of the anemone’s lobe that does not erupt.  The spire-making reconnection simultaneously makes new hot loops that are added to the outside of the lobe that does not erupt.
		     
		          [Throughout this paper, we use the term “jet” to refer to the entire jet event, including the eruption of the minifilament-carrying lobe of the jet-base anemone, the production of the jet bright point by runaway reconnection under the erupting minifilament flux rope inside the erupting lobe, and the production of both the spire and the hot loops added to the non-erupting lobe by the runaway reconnection of the erupting lobe with encountered far-reaching ambient field.  The term “runaway reconnection” refers to reconnection that is driven by the eruption and makes the eruption more explosive by further unleashing it \citep{moore80,moore92,sterling05}.  Following \cite{moore18} we call the base-anemone lobe that does not erupt the jet-base interior, and call brightening in that lobe from the new hot loops added to it by the erupting lobe’s external reconnection base-interior brightening.  We often call that external reconnection “breakout reconnection.”  We use the terms “minifilament eruption” and “jet eruption” as synonyms for the eruption of the whole minifilament-carrying lobe, that is, for the whole magnetic explosion that drives the production of the jet bright point, the base-interior brightening, and the spire via the internal and external reconnection.  We use the term “erupting minifilament” to refer to only the erupting minifilament itself as a tracer of the core of the erupting lobe field.  We take the start of the rise of the minifilament to be the start of the eruption of the minifilament-enveloping lobe, that is, the start of the jet eruption.]
		     
		     In the scenario of  \citep{sterling15}, if the minifilament eruption is a blowout eruption, a so-called blowout jet \citep{moore10} is produced.  The external reconnection opens practically all of the erupting lobe field including the erupting flux-rope core carrying the cool minifilament plasma, resulting in a spire that is comparable in width to the base and has both X-ray-emitting hot plasma and cool minifilament plasma in it.  If the minifilament eruption is a confined eruption, a so-called standard jet  \citep{moore10}  is produced.  The external reconnection opens only a fraction of the erupting lobe field, at most including a small fraction of the field threading the cool minifilament plasma, resulting in a spire that is narrower than in blowout jets and usually does not have appreciable cool (T $<<$ 10$^6$ K) plasma in it \citep{moore13}.
		     
		From subsequent studies of $\sim$  90  randomly selected central-disk coronal jets in quiet regions and coronal holes, using AIA EUV images and magnetograms from SDO’s Helioseismic and Magnetic Imager (HMI), \cite{panesar16b,panesar17,panesar18a} and \cite{mcglasson19}  have shown that at least 90\% of non-active-region coronal jets are made by minifilament eruption as in the scenario of \cite{sterling15}.  In at least 85\%, the magnetograms show persistent convection-driven flux cancelation at the PIL of the pre-eruption minifilament, evidence that this process builds the minifilament-holding flux rope and/or sheared core field and triggers it to erupt.

		     The \cite{sterling15} concept for jet production by minifilament eruption tacitly assumes that the minifilament eruption begins the same way that a larger filament eruption – of the size of those that make CMEs – would begin if at eruption onset the two magnetic fields differed only in scale, not in configuration \citep{sterling18}.  The pre-eruption magnetic field of a jet-making minifilament eruption is a magnetic arcade (an anemone lobe) that has the minifilament-holding flux rope and/or sheared field in its core and has an outside magnetic null or current sheet between it and oppositely-directed far-reaching field.  This set up is essentially the same as for a filament eruption in the middle lobe of a quadrupolar magnetic field: the middle lobe is a magnetic arcade that has a filament-holding flux-rope/sheared-field core and has an outside magnetic null or current sheet between it and the quadrupole’s oppositely-directed overarching field \citep[e.g.][]{antiochos99,moore06}.  From reviewing observations and modeling of CME-producing eruptions of filament-cored magnetic arcades in quadrupolar settings, \cite{moore06} concluded that, depending on details of the field arrangement, these eruptions are started in diverse ways by three mechanisms acting singly or in any combination: (1) runaway internal tether-cutting reconnection under the filament flux rope, (2) runaway breakout reconnection at the external current sheet, and (3) ideal MHD instability, i.e., eruptive instability that does not involve runaway reconnection.
		     
		     If jet-making minifilament eruptions are indeed miniature versions of larger filament eruptions that have access to breakout reconnection, including those that make CMEs, then it should be expected that they display the diversity of onsets displayed by CME-producing filament eruptions that have access to breakout reconnection.  To test this expectation,  \cite{moore18} studied the onsets of 15 of the 20 near-limb polar X-ray jets studied by \cite{sterling15}, the 15 for which the onset of the jet bright point was not hidden behind the limb.  For each jet they compared three times: (1) the time of first detection of the jet bright point in XRT images, that is, the time of the first detectable signature of the runaway internal reconnection, (2) the time of first detection of base-interior brightening, that is, the time of the first detectable signature of the breakout reconnection\footnote{In each of the 15 jets studied by \cite{moore18}, of the base-interior brightening and the spire, the two signatures of the runaway external reconnection (breakout reconnection), the base-interior brightening's first detection was either before or simultaneous with the spire’s first detection.}, and (3) the time of first detection of the rise of the erupting minifilament.  Observing the minifilament to be rising before the onset of the jet bright point and before the onset of the base-interior brightening would indicate that the eruption was initiated by an MHD instability not involving runaway reconnection.  Observing that either the jet bright point or the base-interior brightening started before the first detection of the minifilament’s rise would suggest that the eruption was initiated by an MHD instability involving runaway reconnection, but would still allow the runaway reconnection to have started together with or even after the start of the minifilament’s rise, because the minifilament’s rise might have been hidden from view until the minifilament rose above the obscuring spicule forest of the foreground chromosphere and cooler (T $< $10$^5$ K) transition region.

		     For these 15 jets, in agreement with the idea that jet-making minifilament eruptions are miniature versions of larger filament eruptions that have access to breakout reconnection, the temporal ordering of the first detections of the jet bright point, the base-interior brightening, and the minifilament rise did display a wide diversity, similar to the diversity of the ordering of the onsets of brightening signatures of runaway internal and breakout reconnection relative to the start of the filament’s slow rise in filament eruptions that have access to breakout reconnection and make CMEs.  From their observations, \cite{moore18} surmised that gradual evolution of the jet-base anemone, driven by photospheric convection, often builds an appreciable current sheet between the anemone’s eruptive lobe and the oppositely-directed ambient far-reaching field before the minifilament starts rising, and that runaway breakout reconnection at that current sheet then either starts the eruption or sets in early in the slow-rise onset of the minifilament eruption.

		The present paper follows up on \cite{moore18} by examining 10 randomly-selected coronal jets observed in quiet regions on the central disk, no more than about 30 heliocentric degrees from disk center.  This allows us to see more clearly than in the jets studied by \cite{moore18} – each of which was near the limb – when the minifilament begins to rise relative to when the jet bright point and base-interior brightening each begin.  The rise of the minifilament in its eruption in each of the central-disk jets, as is typical of larger filament eruptions \citep{moore06}, begins with a stepped or smooth slow rise of uneven, constant, or steadily increasing speed before it rapidly accelerates to a much faster speed.  We measured the proper-motion speed (the speed in the plane of the sky) of each erupting minifilament at the beginning and at the end of its slow rise.  The proper-motion speed is a lower bound on the 3D velocity of the erupting minifilament.    In 7 of the 10 jet-eruption onsets, the jet bright point and the base-interior brightening start after the minifilament starts rising.  In agreement with \cite{moore18}, this indicates that in most quiet-region and coronal-hole jets the eruption is initiated by an eruptive MHD instability that does not involve runaway reconnection at first but soon leads to runaway breakout and internal reconnection.  In each of the 10 eruption onsets, the base-interior brightening starts when the minifilament is already rising with a proper-motion speed of about 1 \kms\ or more.  From this, and the fortuitously observed development of the external current sheet in one minifilament eruption, contrary to the surmise of \cite{moore18}, we surmise that a current sheet of appreciable extent seldom develops between the pre-eruption minifilament-enveloping lobe of the jet-base anemone and the ambient far-reaching field before the start of the eruption because that lobe’s pre-eruption evolutionary upwelling is seldom fast enough to build up an appreciable current sheet there faster than reconnection tears it down.
	
	 \vspace{-0.7cm}
	\section{DATA SET}\label{data} 
	     For each of our coronal-jet eruptions, the date, location on the central disk, and minifilament eruption start time are given in Table \ref{tab:list}.  These jets are the 10 randomly selected quiet-region jets previously studied by \cite{panesar16b,panesar17}\footnote{The ten jets studied in \cite{panesar16b} were randomly selected as follows.  We looked for any quiet-region obvious jet that showed a bright spire in JHelioviewer AIA 171 \AA\ full-disk images taken during 2012.  This was not intended to be a comprehensive search, as we used only a coarse time cadence – 30 min, which is longer than the ~10-min lifetime of many X-ray polar-coronal-hole jets (e.g. \citealt{savcheva07}). While our selected jets occurred at random times, our search did favor larger, easy-to-identify, outstanding examples. After collecting ten obvious jets, we stopped looking for more, studied those ten jets, and reported our findings in \cite{panesar16b}.}, and are numbered (J1-J10) and listed in Table \ref{tab:list} in the same order as in  \cite{panesar16b}.  These jets were found by  \cite{panesar16b} by viewing full-disk AIA 171 \AA\ movies, using JHelioviewer \citep{muller17}.  They also viewed each jet in other AIA EUV channels and found that before and during its eruption the minifilament was most clearly seen in the 171 \AA\ images.  [AIA images have 0\arcsec.6 pixel$^{-1}$ and 12 s cadence \citep{lem12}.]  From AIA 171 \AA\ images and HMI magnetograms,  \cite{panesar16b,panesar17} found that each jet was made by a minifilament eruption, and that the sheared-field/flux-rope core field carrying the minifilament was evidently built and triggered to erupt by persistent flux cancelation at the PIL under the minifilament.  [HMI line-of-sight magnetograms have 0\arcsec.5 pixel$^{-1}$, 45 s cadence, and a noise level of about 10 G \citep{schou12,scherrer12}.]  In its AIA 171 \AA\ images, each of these jets is a wide-spire jet, apparently made by blowout eruption of the minifilament and its enveloping magnetic field.  For the present study, we used the AIA 171 \AA\  full-cadence (12 s) movies of  \cite{panesar16b} to closely examine the onset of each jet eruption. AIA and HMI data sets are analyzed using SSW routines \citep{freeland98}.

\begin{figure*}
	\centering
	\includegraphics[width=\linewidth]{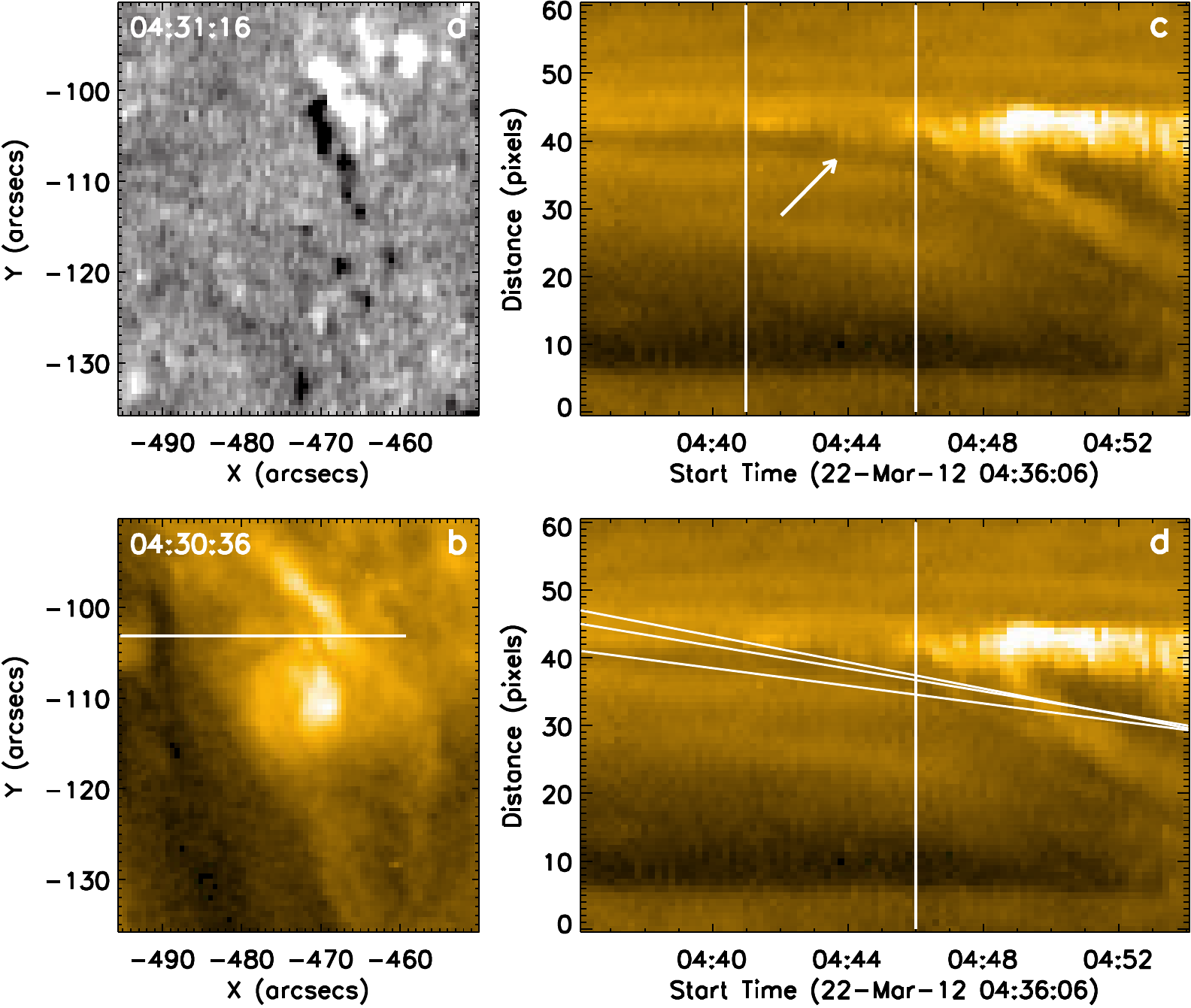}
	\caption{Magnetic setting and intensity time-distance map of the minifilament in the eruption of jet J1.  Panels (a) and (b) respectively show an HMI magnetogram and an AIA 171 \AA\ image of the jet base and pre-jet minifilament. The white line in (b) shows the cut for the intensity time-distance map shown in (c).  In (c), the arrow points to the dark track of the minifilament and the two vertical lines mark the start time (04:41:00 UT) and end time (04:46:00 UT) of the minifilament’s slow rise.  Panel (d) shows the same time-distance plot of panel (c) but with the three tangent lines that were used to measure the proper-motion speed of the minifilament at the end of its slow rise/start of its fast rise. HMI contours (level $\pm$50 G) of panel (a)  are overlaid on panel (b), where turquoise and red,
		represent positive and negative polarities, respectively. North is upward and west is to the right in all solar images in this paper.
	} \label{fig1a}
\end{figure*} 	  

	\begin{figure*}
		\centering
		\includegraphics[width=\linewidth]{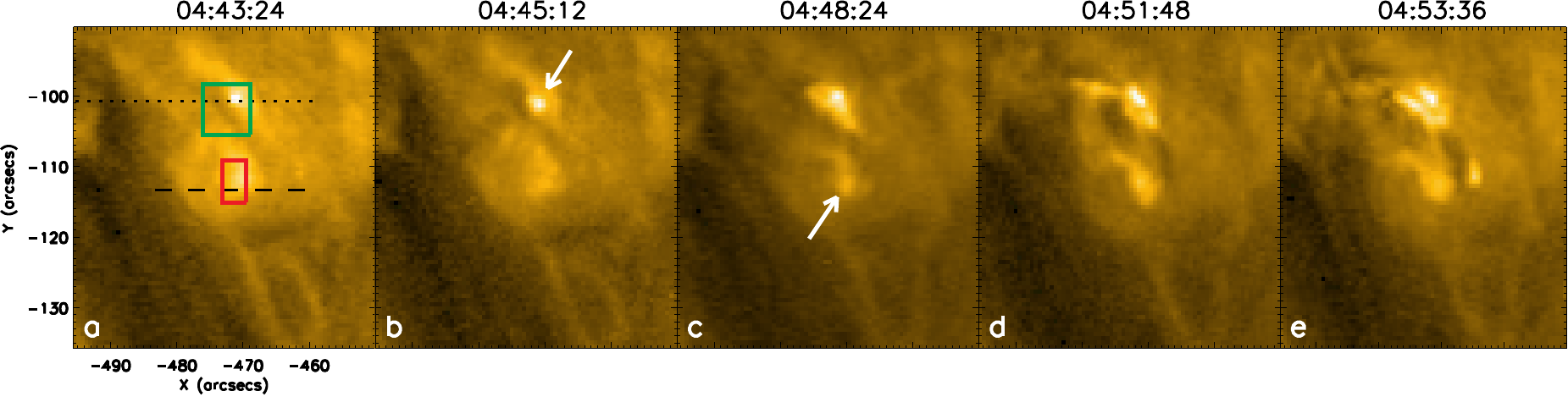}
		\caption{AIA 171 \AA\ images of the eruption of the minifilament and onset and growth of the jet bright point and base-interior brightening in the base of jet J1.  The green and red boxes in (a) outline the areas in which the area-average jet bright point intensity and base-interior brightening intensity were measured for the corresponding intensity-time plots in Figures \ref{fig1c}a and \ref{fig1c}c. The dotted and dashed lines respectively mark the cuts for the intensity time-distance maps shown in Figures \ref{fig1c}b and \ref{fig1c}d. The arrow in (b) points to the growing jet bright point under the rising minifilament.  The arrow in (c) points to the base-interior brightening early in its onset.  MOVIE1 is an animation of this Figure.  } \label{fig1b}
	\end{figure*}

	     From the movie of each jet, we visually identified the pre-jet minifilament, the slow-rise and fast-rise phases of the erupting minifilament, the jet bright point, the onset of the base-interior brightening, and the onset of the spire.  By stepping through the movie frame by frame, we identified the time of the start of the minifilament’s slow rise, the time of the end of the minifilament’s slow rise, and the time of first detection of the jet bright point, the base-interior brightening, and the spire, and visually estimated the uncertainty in each of these five times.  These times and their uncertainties are listed in Table \ref{tab:list}.
	     
	The start time of the minifilament’s slow rise is the time of the first frame of the AIA 171 \AA\ movie in which we could visually discern that the minifilament had started rising.  We take the end time of the minifilament’s slow rise to also be the start time of the minifilament’s fast rise.  It is the time of the movie frame in which we judged the rising minifilament to be midway through the brief high-acceleration transition from the slow-rise phase to the fast-rise phase of its eruption.  By this time, the minifilament is and has been undergoing greater acceleration than immediately earlier in its slow rise.  Therefore, for each of our 10 erupting minifilaments, because the rising minifilament does not obviously slow down before the start of the slow-rise/fast-rise transition, we expect the measured rise speed of the minifilament at the slow-rise end time to be faster than at the slow-rise start time.
	
	     For each erupting minifilament, from viewing the movie we selected a straight line (as for jet J1 in Figure \ref{fig1a}) along which the time-distance plot of the 171 \AA\ intensity gave a good track of the minifilament from before the onset through the slow-rise phase and into the fast-rise phase of the eruption.  To measure the proper-motion speed of a rising minifilament at the start of the slow rise, we visually drew three tangent lines to the time-distance plot’s dark trace of the minifilament at the slow-rise start time, each line drawn separately in the absence of the other two lines.  One line was drawn tangent to the upper edge of the trace, a second was drawn tangent to the lower edge of the trace, and the third was drawn tangent to the middle of the trace at the start time.  The slope of each tangent gives an estimate of the of the proper-motion speed of the minifilament at slow-rise onset.  For the measured speed, we took the average of the three speeds given by the three tangents, and for the uncertainty in the measured speed we took the square root of the mean of the squares of the differences of the three speeds from their average.  In the same way, we measured the proper-motion speed of the rising minifilament at the end of the slow rise.  For jet J1, the three tangent lines from which the proper-motion speed of the minifilament was measured at the end of the minifilament’s slow rise are shown in Figure \ref{fig1a}.  For each of the 10 jet eruptions, the minifilament’s measured speed and speed uncertainty at the start and at the end of the minifilament’s slow rise are given in Table \ref{tab:list}.
	     
	
	\begin{figure*}
		\centering
		\includegraphics[width=\linewidth]{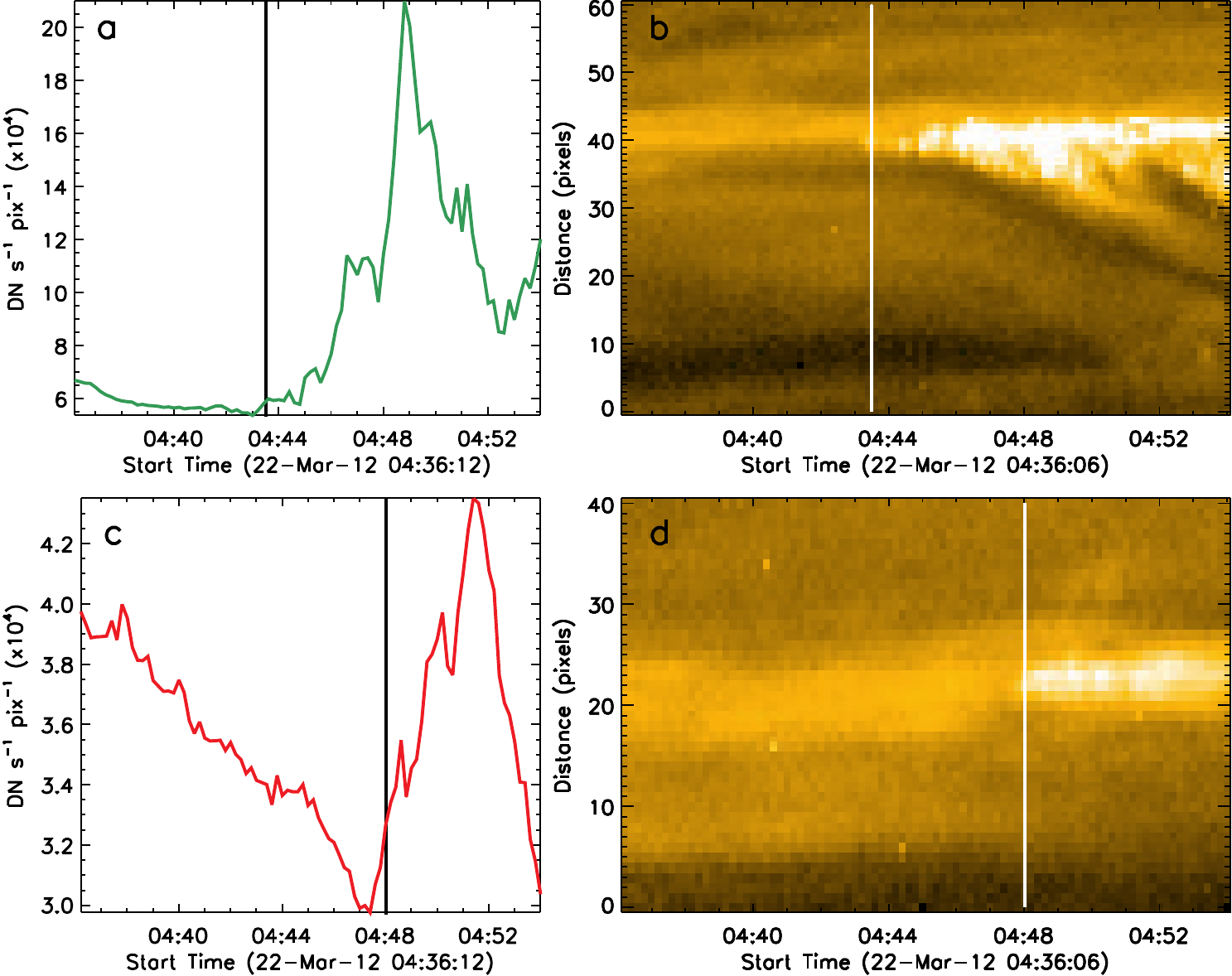}
		\caption{Intensity-time plots and intensity time-distance maps for the jet bright point and base-interior brightening in the base of jet J1.  Panel (a) shows the AIA 171 \AA\ area-average intensity as a function of time for the jet bright point in the green box in Figure \ref{fig1b}.  Panel (c) shows the area-average intensity as a function of time for the base-interior brightening in the red box in Figure \ref{fig1b}.  Panels (b) and (d) respectively show the AIA 171 \AA\ intensity time-distance maps from the dotted and dashed lines through the green and red boxes in Figure {fig1b}.    The jet bright point start time is marked by a vertical line in panels (a) and (b).  The base-interior brightening start time is marked by a vertical line in panels (c) and (d).} \label{fig1c}
	\end{figure*} 
	\section{Results}\label{result}
	Most of our results are from the start times and minifilament slow-rise speeds listed in Table \ref{tab:list} for our 10 jet eruption onsets.  Before addressing the results from Table \ref{tab:list} as a whole, we present the observations in detail for the eruption onsets of three example jets: J1, J7, and J9 in Table \ref{tab:list}.
	
	\subsection{\textit{Jet J1 }\label{jet1}}
	          Jet eruption J1 is an example of an eruption in which the minifilament’s rise started first, the jet bright point started second, the base-interior brightening started third, and the spire started last.   This timing indicates that the minifilament eruption began by an MHD eruptive instability not involving runaway reconnection and was later made more explosive by runaway internal tether-cutting reconnection and breakout reconnection.
	          
	Figure \ref{fig1a}(b) is an AIA 171 \AA\ image showing the minifilament in the jet’s base 10 minutes before the start of the minifilament’s slow rise.  The co-temporal HMI magnetogram in panel (a) shows the underlying PIL at which flux cancelation evidently built the minifilament-holding field and triggered it to erupt (see \citealt{panesar17}).  The magnetogram and the 171 \AA\ image together indicate that the base of jet J1 was a magnetic anemone centered on an island of negative-polarity flux in a region of far-reaching field rooted in positive-polarity flux.  A lobe of the anemone enveloped the minifilament and its flux-cancelation PIL.  Some of the rest of the anemone, i.e., some of the base interior, is presumably the bright blob on the southeast side of the minifilament in the 171 \AA\ image.  This blob is presumably a signature of the base-interior anemone lobe that connects negative flux east/southeast to positive flux that the magnetogram barely detects.  The minifilament eruption blows out to the east into ambient field that reaches far south (see MOVIE1).

     The horizontal white line in Figure \ref{fig1a}(b) is the cross-minifilament cut along which the AIA 171 \AA\ intensity time-distance plot in panel (c) was made.  The time-distance plot starts 5 minutes before the start of the minifilament’s slow rise and ends well into the fast rise.  The arrow in panel (c) points to the dark track of the minifilament.  As is normal for filament and minifilament eruptions, the track shows a sharp bend, a sharp increase in rise speed, at the end time of the slow rise (by definition, the end time of the slow rise is the start time of the fast rise).  The two vertical lines in panel (c) mark the start time (04:41:00 UT) and end time (04:46:00 UT) of the slow rise, which times were found visually by stepping through the AIA 171 \AA\ movie and, to within their uncertainties (Table \ref{tab:list}), coincide with the start and end of the slow-rise interval of the minifilament’s time-distance track.  Figure \ref{fig1a}d shows the time-distance plot of panel (c) with the end time (04:46:00 UT) of the slow rise marked by the vertical line.  The three sloped lines in panel (d), drawn tangent to the top, middle, and bottom of the minifilament’s track at the end time of the slow rise, are the lines by which the proper-motion speed of the minifilament at the end of its slow rise was measured and found to be 6.0$\pm$1.0 \kms.  In the same way, the minifilament’s rise speed at the slow-rise start time was measured to be 1.2$\pm$0.3 \kms, showing that, as expected, the slow-rise speed of this erupting minifilament was measurably faster at the end of its slow rise than at the start.
	\begin{figure*}
		\centering
		\includegraphics[width=\linewidth]{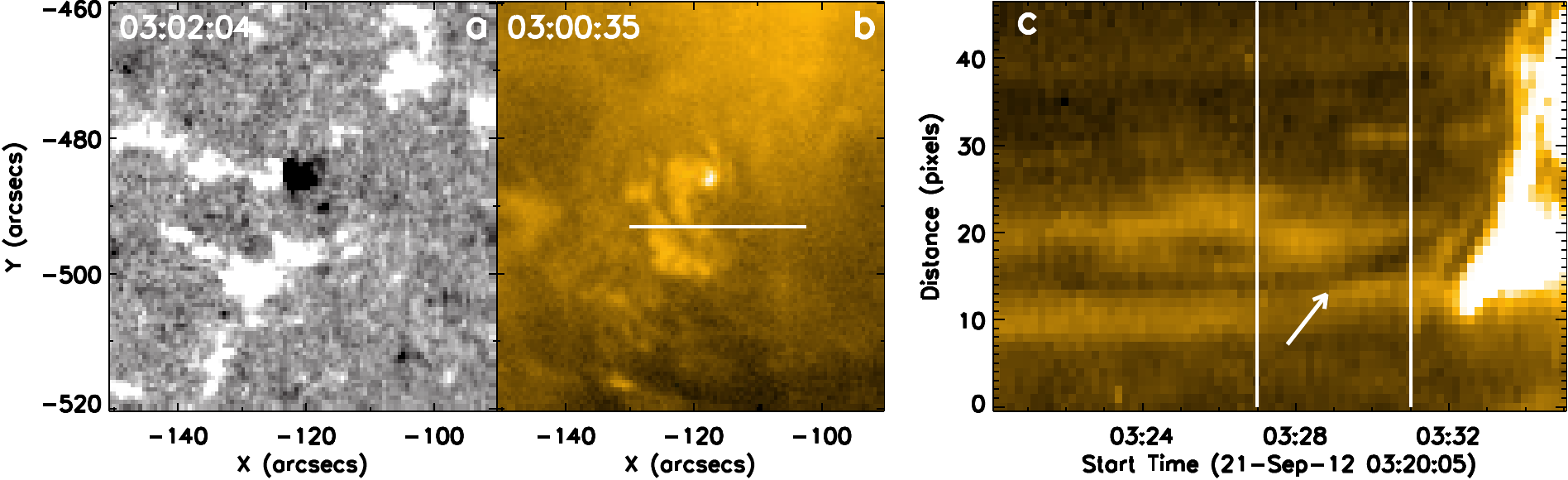}
		\caption{Magnetic setting and intensity time-distance map of the minifilament in the eruption of jet J7.  Panels (a) and (b) respectively show an HMI magnetogram and an AIA 171 \AA\ image of the jet base and pre-jet minifilament.  The white line in (b) shows the cut for the intensity time-distance map shown in (c).  In (c), the arrow points to the dark track of the minifilament and the two vertical lines mark the start time (03:27:11 UT) and end time (03:31:00 UT) of the minifilament’s slow rise. HMI contours (level $\pm$50 G) of panel (a)  are overlaid on panel (b), where turquoise and red, represent positive and negative polarities, respectively.} \label{fig2a} 
	\end{figure*} 
	
	
	Figure \ref{fig1b} is a sequence of AIA 171 \AA\ images showing the development of the jet bright point and the base-interior brightening as the minifilament erupts.  The image in panel (a) is at the start time (04:43:24 UT) of the jet bright point on the PIL under the slowly rising minifilament.  The arrow in panel (b) points to the growing jet bright point two minutes later.   Panel (c) shows the minifilament about two minutes into its fast rise.  The arrow in panel (c) points to the base-interior brightening early in its onset.  Panels (d) and (e) show further development of the jet bright point and base-interior brightening as the blowout eruption of the minifilament progresses.
	
	Figure \ref{fig1c}(a) is the time plot of the average AIA 171 \AA\ intensity per pixel in the green outlined box in panel (a) of Figure \ref{fig1b}.  The box covers the jet bright point.  The intensity-time plot confirms the jet bright point’s start time (04:43:24 $\pm$ 12 s UT) that was found visually by stepping through the 12-second-cadence 171 \AA\ movie.  That time is marked by the vertical line in intensity-time plot.  Figure \ref{fig1c}(b)  is the AIA 171 \AA\ intensity time-distance plot from the horizontal line through the jet bright point in Figure \ref{fig1b}(a).  This plot also confirms that the jet bright point started at 04:43:24 $\pm$ 12 s, which time is marked in this plot by the vertical line.  Thus, the plots in panels (a) and (b) confirm that the jet bright point in jet J1 started at least a minute after the minifilament started rising at 04:41:00 $\pm$ 60 s.

	Figure \ref{fig1c}(c) is the time plot of the average 171 \AA\ intensity per pixel in the red outlined box in panel (a) of Figure \ref{fig1b}.  The box covers the location of the onset of the base-interior brightening.  The intensity time plot confirms the base-interior brightening’s start time (04:48:00$\pm$24 s) that was found visually by stepping through the 171 \AA\ movie.  That time is marked by the vertical line in Figure \ref{fig1c}(c).  Figure \ref{fig1c}(d) is the AIA 171 \AA\ intensity time-distance plot from the horizontal line through the red box in Figure \ref{fig1b}a.  That line runs through the location of the onset of the base-interior brightening.  This plot also confirms that the base-interior brightening started at 04:48:00$\pm$24 s, which time is marked by the vertical line in panel (d).   
	     
	Thus, the plots in Figure \ref{fig1c} confirm that the base-interior brightening in jet J1 started about four minutes after the jet bright point started and about six minutes after the minifilament’s slow rise started.  This timing is evidence that the eruption that produced J1 started by an ideal MHD eruptive instability that at first did not involve either internal or breakout runaway reconnection.

	\subsection{\textit{Jet J7 }\label{jet2}}
	
	The eruption of jet J7 is an example of a jet eruption in which, within the uncertainty in the start times, the rising of the minifilament and the brightening of the jet bright point started together, the first detection of the base-interior brightening was later, and the first detection of the spire was later yet.  This timing allows the instability that initiated the eruption to have involved, at the very start of the eruption, the runaway internal tether-cutting reconnection that made the jet bright point, and indicates that the breakout reconnection that made the base-interior brightening and the spire set in after the eruption had started.
	
\begin{figure*}
		\centering
		\includegraphics[width=\linewidth]{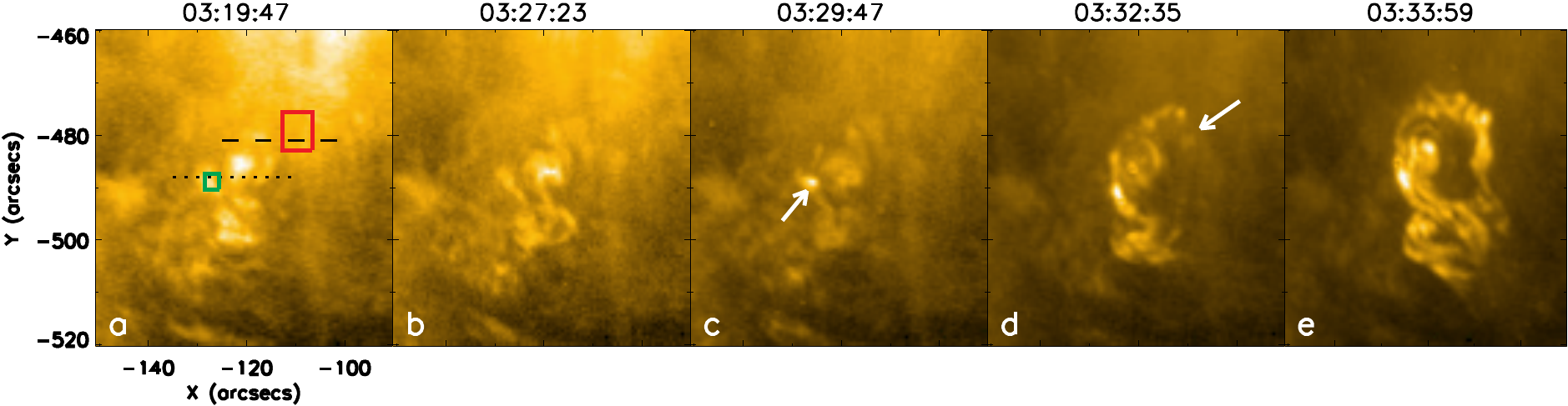}
		\caption{AIA 171 \AA\ images of the eruption of the minifilament and onset and growth of the jet bright point and base-interior brightening in the base of jet J7.  The green and red boxes in (a) outline the areas in which the area-average jet bright point intensity and base-interior brightening intensity were measured for the corresponding intensity-time plots in Figure \ref{fig2c}a and \ref{fig2c}c.  The dotted and dashed lines respectively show the cuts for the intensity time-distance maps shown in Figure \ref{fig2c}b and Figure \ref{fig2c}d.  The arrow in (c) points to the growing jet bright point under the rising minifilament.  The arrow in (d) points to the base-interior brightening early in its onset.  MOVIE2 is an animation of this Figure.
		} \label{fig2b} 
	\end{figure*}

	Figure \ref{fig2a}(b) is an AIA 171 \AA\ image showing the minifilament in the jet’s base half an hour before the minifilament’s slow rise.  The co-temporal HMI magnetogram in panel (a) shows the underlying flux-cancelation PIL traced by the minifilament.  The magnetogram in Figure \ref{fig2a}, the placement of the jet bright point and base-interior brightening seen in Figure \ref{fig2b}, and the direction of the blowout eruption of the minifilament seen in Figure \ref{fig2b} indicate that the base of jet J7 was a magnetic anemone centered on an island of negative-polarity flux in a region of magnetic field that was rooted in positive-polarity flux and reached far to the southwest (see MOVIE2).

	     The horizontal white line in Figure \ref{fig2a}(b) is the minifilament-crossing cut along which the AIA 171 \AA\ intensity time-distance plot in panel (c) was made.  The time-distance plot begins well before the start of the minifilament’s slow rise and ends well into the fast rise.  The arrow in panel (c) points to the dark track of the minifilament.  The two vertical lines in panel (c) mark the start time (03:27:11 UT) and end time (03:31:00 UT) of the minifilament’s slow rise.  The slope of the minifilament’s dark track in the time-distance plot obviously increases through the slow rise and is suddenly steeper immediately after the end of the slow rise, early in the fast rise.  The measured proper-motion speed of the rising minifilament is 3.0 $\pm$ 0.5 \kms\ at the start of the slow rise and 17.0 $\pm$ 1.0 km s-1 at the end of the slow rise (Table \ref{tab:list}).
	
		Figure \ref{fig2b} is a sequence of five AIA 171 \AA\ images of the jet base from several minutes before eruption onset to a few minutes into the fast rise of the minifilament’s blowout eruption.  Panel (b) shows the minifilament at 03:27:23 UT, which is within the uncertainty range of the minifilament’s slow-rise start time (03:27:11 $\pm$ 90 s) and about a minute before the jet bright point was first detected at 03:28:11 $\pm$ 24 s.  The arrow in panel (c) points to the growing jet bright point about a minute and a half after its start time.  The arrow in panel (d) points to a growing part of the base-interior brightening that is much bigger and brighter in panel (e), a minute and a half later.
	
\begin{figure*}
		\centering
		\includegraphics[width=\linewidth]{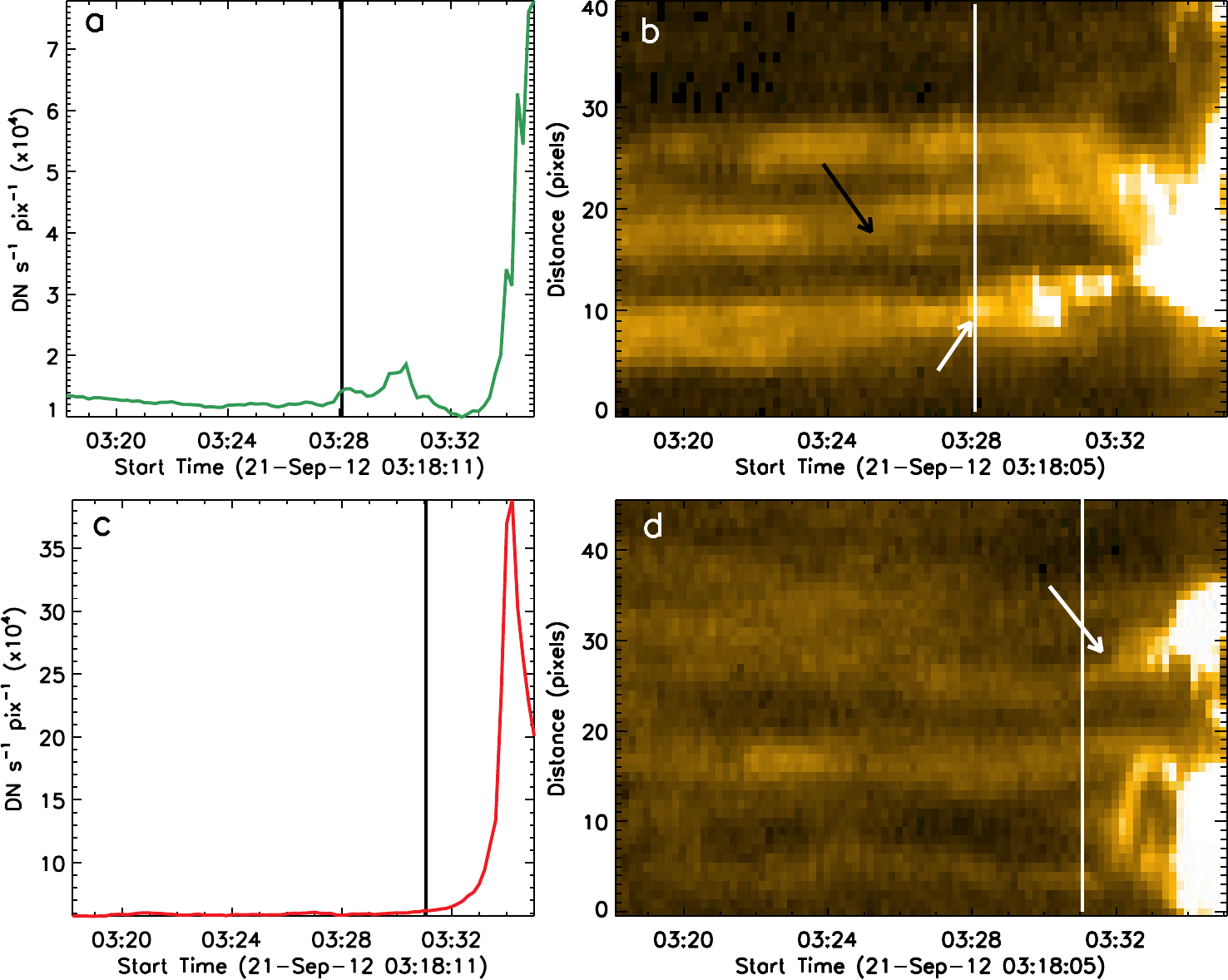}
		\caption{Intensity-time plots and intensity time-distance maps for the jet bright point and base-interior brightening in the base of jet J7.  Panel (a) shows the AIA 171 Å average intensity as a function of time for the jet bright point in the green box in Figure \ref{fig2b}a.  Panel (c) shows the AIA 171 \AA\ area-average intensity as a function of time for the base-interior brightening in the red box in Figure \ref{fig2b}.  Panels (b) and (d) respectively show the AIA 171 \AA\ intensity time-distance maps from the dotted and dashed lines through the green and red boxes in \ref{fig2b}.  The black and white arrows in (b) respectively point to the tracks of the minifilament and the jet bright point.  The white arrow in (d) points to the track of the growing base-interior brightening.  The jet bright point start time is marked by a vertical line in panels (a) and (b).  The base-interior brightening start time is marked by a vertical line in panels (c) and (d).} \label{fig2c}
	\end{figure*}

	     Figure \ref{fig2c}(a) is the time plot of the AIA 171 \AA\ average intensity per pixel in the green outlined box in panel (a) of Figure \ref{fig2b}.  That box is centered on the spot where the jet bright point starts.  The vertical line in Figure \ref{fig2c}(a) marks the jet bright point’s start time (03:28:11 $\pm$ 24 s) that was found visually by stepping through the 12-second-cadence AIA 171 \AA\ movie.  The intensity-time plot confirms that the jet bright point started then.  In panel (a), the intensity goes through a minimum at about 03:32:30 UT, early in the minifilament’s fast rise, because, in its blowout eruption, the dark minifilament passes in front of the jet bright point (see MOVIE2).  Figure \ref{fig2c}(b) is the AIA 171 \AA\ intensity time-distance plot from the horizontal line through the green box in Figure \ref{fig2b}(a).  The black arrow points to the dark track of the minifilament, and the white arrow points to the onset of the jet bright point’s brightening.  The vertical line in panel (b) again marks the jet bright point’s start time found visually from the AIA 171 \AA\ movie.  Thus, panel (b) confirms that the jet bright point in jet J7 started within the uncertainty of the minifilament’s slow-rise start time of 03:27:11 $\pm$ 90 s.

	Figure \ref{fig2c}(c) is the time plot of the AIA 171 \AA\  average intensity per pixel in the red outlined box in panel (a) of Figure  \ref{fig2b}.  The box covers the part of the base-interior brightening from which the base-interior brightening start time (03:31:11 $\pm$ 12 s) was visually found from the AIA 171 \AA\ movie.  The vertical line marks 03:31:11 UT, showing that the base-interior brightening intensity-time plot is in agreement with 03:31:11 $\pm$ 12 s being the base-interior brightening start time.  Figure \ref{fig2c}(d) is the AIA 171 \AA\ intensity time-distance plot from the base-interior-brightening-crossing horizontal line through the red box in Figure \ref{fig2b}.  The arrow in the time-distance plot points to the base-interior brightening’s track early in the base-interior brightening’s brightening.  The vertical line marks 03:31:11 UT, again verifying the base-interior brightening’s start time of 03:31:11 $\pm$ 12 s visually found from the AIA 171\AA\ movie.
	
	Thus, the plots in Figure \ref{fig2c} confirm that the jet bright point and the minifilament’s slow rise started together within the  $\pm$ 90 s uncertainty of the visually-found slow-rise start time and the base-interior brightening started a few minutes later.  This timing indicates that runaway internal tether-cutting reconnection helped start the eruption that made jet J7, and that runaway breakout reconnection did not set in until the start of the eruption’s fast rise (Table \ref{tab:list}).


	\subsection{\textit{Jet J9 }\label{jet3}}

	          In the eruption of jet J9, the slow rise of the minifilament and the brightening of the base-interior brightening started together, the spire started next, and the jet bright point started last, at the end of the slow rise/start of the fast rise.  This timing indicates that the instability that started the eruption involved, from the very start of the eruption, the runaway breakout reconnection that made the base-interior brightening and the spire, and that the runaway internal tether-cutting reconnection that made the jet bright point did not start until after the eruption was well underway.  
	          
	          The eruption of J9 is especially instructive because what is apparently the external current sheet for the reconnection that makes the base-interior brightening and the spire is seen in the AIA 171 \AA\ movie.  The feature we take to be the current sheet is a thin bright layer on the erupting minifilament’s flat front, viewed nearly edge-on.  The bright layer grows in extent together with the flat part of the minifilament’s front, both of which appear and start growing late in the minifilament’s slow rise.  This suggests that, when the base-interior brightening-making-breakout reconnection started, the breakout-reconnection current sheet was too small and dim to be seen, and grew large enough and bright enough to be seen only when, as the eruption sped up, the building up of the current sheet became enough faster than the tearing down of the current sheet by the reconnection.
	    
\begin{figure*}
		\centering
		\includegraphics[width=\linewidth]{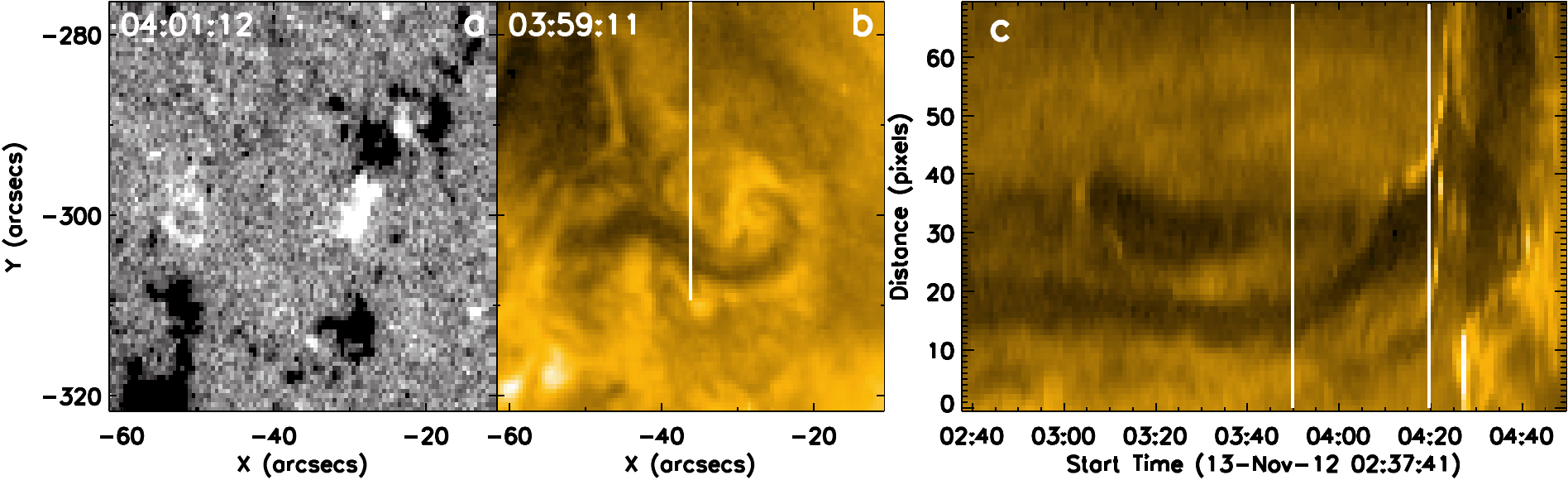} 
		\caption{Magnetic setting and intensity time-distance map of the minifilament in the eruption of jet J9.  Panels (a) and (b) respectively show an HMI magnetogram and an AIA 171 \AA\ image of the jet base and at about 10-minutes into the  minifilament’s 30-minute slow rise.  The white line in (b) shows the cut for the intensity time-distance map shown in (c).  In (c), the two vertical lines mark the start time (03:50:35 UT) and end time (04:20:00 UT) of the minifilament’s slow rise. HMI contours (level $\pm$50 G) of panel (a)  are overlaid on panel (b), where turquoise and red,
				represent positive and negative polarities, respectively.} \label{fig3a}
	\end{figure*}

\begin{figure*}
		\centering
		\includegraphics[width=\linewidth]{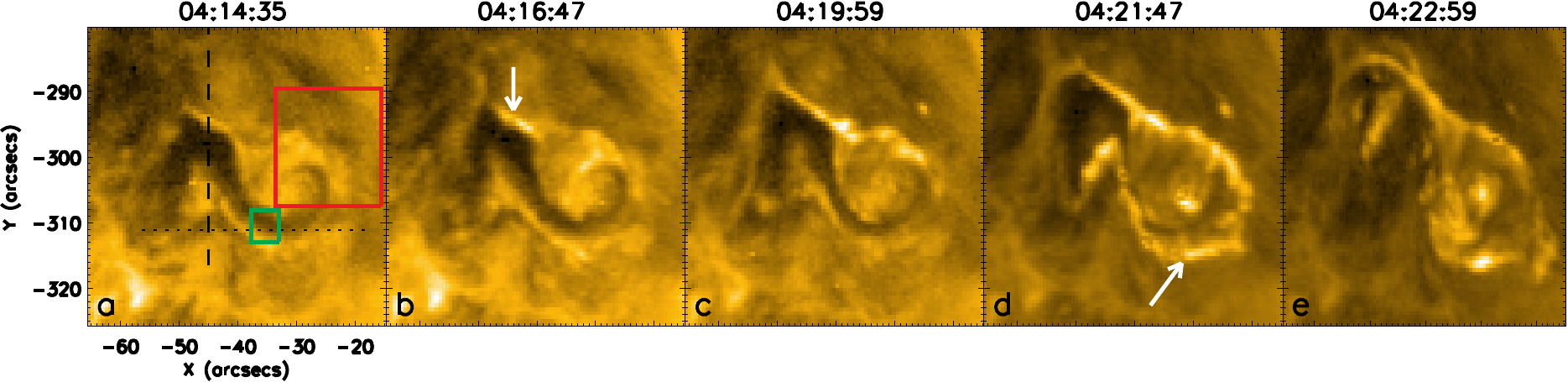} 
		\caption{AIA 171 \AA\ images of the erupting minifilament, onset and growth of the jet bright point, growth of the base-interior brightening, and growth and disruption of the breakout-reconnection current sheet late in the slow rise and early in the fast rise of the eruption of jet J9.  The green and red boxes in (a) outline the areas in which the area-average jet bright point intensity and base-interior brightening intensity were measured for the corresponding intensity-time plots in Figures \ref{fig3c}a and \ref{fig3c}c.  The dotted and dashed lines respectively mark the cuts for the intensity time-distance maps shown in Figures \ref{fig3c}b and \ref{fig3c}d.  The arrow in (b) points to the breakout-reconnection current sheet.  The arrow in (d) points to the onset of the jet bright point.  MOVIE3 is an animation of this Figure.  
	} \label{fig3b}
	\end{figure*}

	         Figure \ref{fig3a}(b)  is an AIA 171 \AA\ image showing the minifilament about 10 minutes into the  minifilament’s 30-minute slow rise.  The co-temporal HMI magnetogram in panel (a) shows that the flux-cancelation PIL traced by the minifilament is between positive-polarity minority flux and surrounding negative-polarity majority flux.  The magnetogram, the placement of the jet bright point and base-interior brightening seen in Figure \ref{fig3b}, and the direction of the minifilament’s blowout eruption seen in Figure \ref{fig3b} are all consistent with the base of jet J9 being a magnetic anemone centered on the island of positive flux in a region of negative flux in the feet of magnetic field reaching far to the north (see MOVIE3).
	
	     The vertical white line in Figure \ref{fig3a}(b) is the minifilament-crossing cut from which the AIA 171  \AA\ intensity time-distance plot in panel (c) was made.  The time-distance plot runs from over an hour before the start of the minifilment’s slow rise to half an hour after the start of the minifilament’s fast rise.  The two vertical lines in panel (c) mark the start time (03:50:35 UT) and the end time (04:20:00 UT) of the minifilament’s slow rise.  The erupting minifilament’s proper-motion speed measured from the minifilament’s track in the time-distance plot in panel (c) is 1.2 $\pm$ 0.5 \kms\ at the start of the slow rise and 4.5 $\pm$ 0.5 \kms\ at the end of the slow rise.  The slope of the minifilament’s track in panel (c) is roughly constant through the slow rise, but is perhaps a little steeper in the second half of the slow rise than in the first half.

	     Figure \ref{fig3b}  is a sequence of five AIA 171 \AA\ images of the jet base spanning the last five minutes of the 30-minute slow rise and the first three minutes of the fast rise of the minifilament’s blowout eruption.  In panel (a), at 04:14:35 UT, the minifilament already has the flat, bright, leading-edge feature that we take to be the current sheet between the jet-base anemone’s erupting lobe and encountered ambient far-reaching field.  In MOVIE3 it is seen that earlier in the slow rise the minifilament’s leading edge is rounded and dark.  The leading edge starts to flatten and brighten about a minute before the image in panel (a), and then grows rapidly in length and brightness.  In panel (b), two minutes after panel (a), the arrow points to the flat, bright front, which is not much longer than in panel (a) but is now much brighter.  In panel (c), at the end of the minifilament’s slow rise, three minutes after panel (b), the current-sheet bright front is about twice longer than in panel (b).  In panel (d), two minutes into the fast rise, the current sheet still has about the same length as in panel (c).  In panel (e), a minute later, the current sheet and minifilament are breaking up in the blowout eruption’s fast rise.  The arrow in panel (d) points to the onset of the jet bright point.  A minute later, in panel (e), the jet bright point is much brighter.  The five-image sequence in Figure \ref{fig3b}  also shows the growing base-interior brightening in the negative-polarity northern feet of breakout-reconnection-made loops that link that negative flux to the jet-base anemone’s central positive flux.

\begin{figure*}
			\centering
			\includegraphics[width=\linewidth]{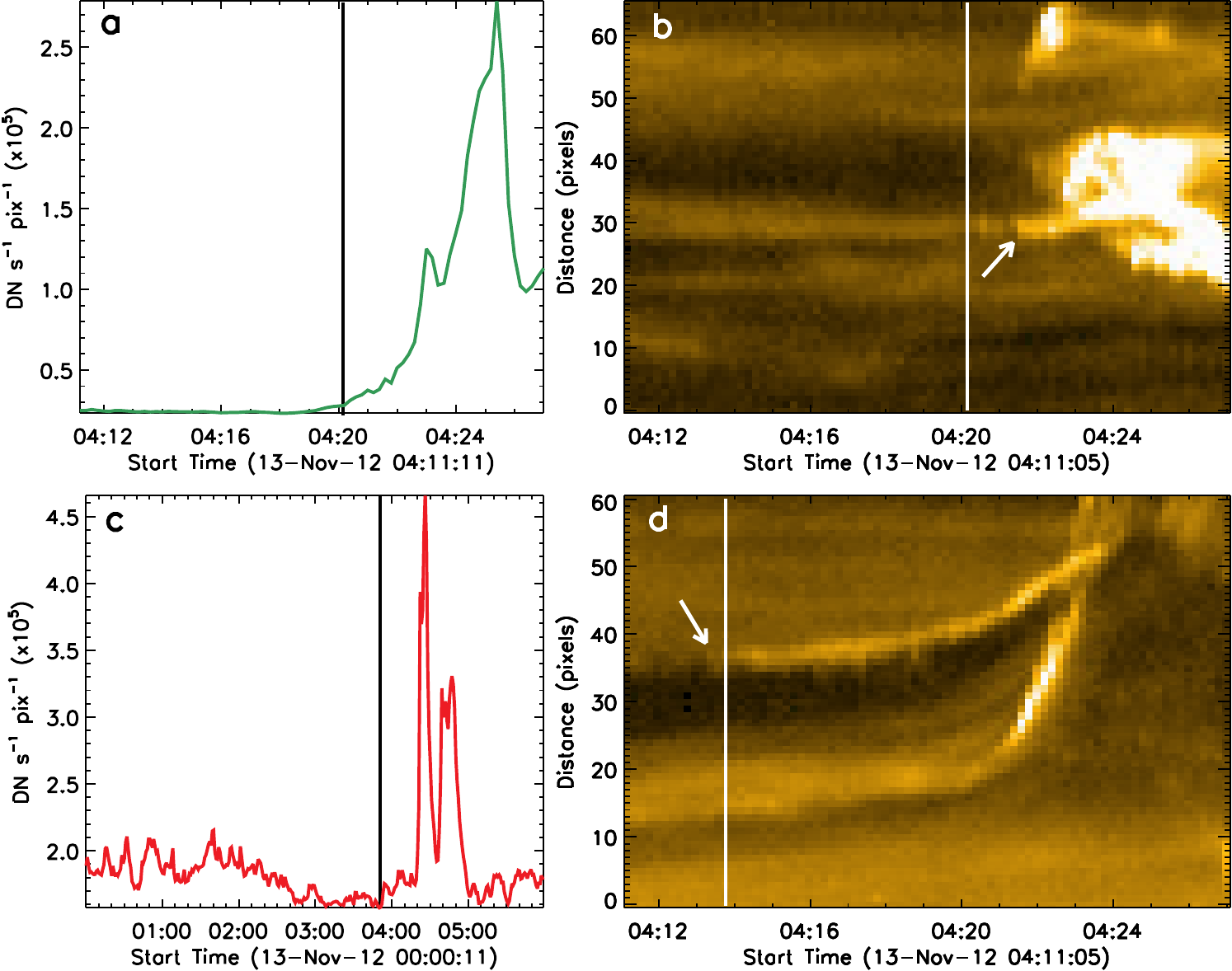}
			\caption{Intensity-time plot and intensity time-distance map for the jet bright point, intensity-time plot for the base-interior brightening, and intensity time-distance map of the breakout-reconnection current sheet in the eruption of jet J9.  Panel (a) shows the AIA 171 \AA\ area-average intensity as a function of time for the jet bright point in the green box in Figure \ref{fig3b}.  Panel (c) shows the AIA 171 \AA\ area-average intensity as function of time for the base-interior brightening in the red box in Figure \ref{fig3b}.  Panels (b) and (d) respectively show the AIA 171 \AA\ intensity time-distance maps from the dotted and dashed lines in Figure \ref{fig3b}.  The jet bright point start time is marked by a vertical line in panels (a) and (b).  The base-interior brightening start time is marked by a vertical line in panel (c).  The time when the breakout current sheet was first discernible in the AIA 171 \AA\ movie is marked by a vertical line in panel (d).} \label{fig3c}
		\end{figure*}

	     Figure \ref{fig3c}(a) is the time plot of the AIA 171 \AA\ average intensity per pixel in the green outlined box in panel (a) of Figure \ref{fig3b}.  The box is centered on where the jet bright point starts.  The vertical line in Figure \ref{fig3c}(a)  marks the jet bright point’s start time (04:20:11 $\pm$24 s) that was estimated visually by stepping through the 12-second-cadence AIA 171 \AA\ movie.  The intensity-time plot in panel (a) confirms that the jet bright point brightening (presumably from runaway internal tether-cutting reconnection of the erupting lobe’s legs under the rising minifilament flux rope) started at about 04:20 UT.  Figure \ref{fig3c}(b) is the AIA 171 \AA\ intensity time-distance plot from the horizontal dotted line through the green box in Figure \ref{fig3b}(a).  The vertical line again marks the jet bright point’s start time (04:20:11 UT) that was found visually by stepping through the AIA 171 \AA\ movie.  The white arrow in panel (b) points to the track of the jet bright point in the time-distance plot.  The jet bright point’s brightening first shows at about 04:20:30 UT, in agreement with the visually estimated jet bright point start time of 04:20:11 $\pm$ 24 s.
	
	Figure \ref{fig3c}(c) is the time plot of the AIA 171 \AA\ average intensity per pixel in the red outlined box in panel (a) of Figure \ref{fig3b}.  The box covers the part of the base-interior brightening from which the base-interior brightening start time (03:50:35$\pm$24 s) was visually found from the AIA 171 \AA\ movie.  The vertical line in Figure \ref{fig3c}(c) marks 03:50:35 UT, showing that the intensity-time plot agrees with 03:50:35$\pm$24 s being the base-interior brightening start time for jet J9.

	     Figure \ref{fig3c}(d) is the AIA 171 \AA\ intensity time-distance plot from the vertical dashed line that cuts across the current-sheet front of the rising minifilament in panel (a) of Figure \ref{fig3b}(a).  The arrow in Figure \ref{fig3c}(d) points to the rising minifilament’s track in this time-distance plot.  The track of the current-sheet bright front of the minifilament shows that the speed of the front increased during the last 6 minutes of the minifilament’s slow rise (04:14 – 04:20 UT).  The vertical line in this plot marks 04:13:47 UT, showing that brightening of the current-sheet front of the rising minifilament first became detectable at that time, late in the minifilament’s slow rise.

\begin{figure}
			\centering
			\includegraphics[width=\linewidth]{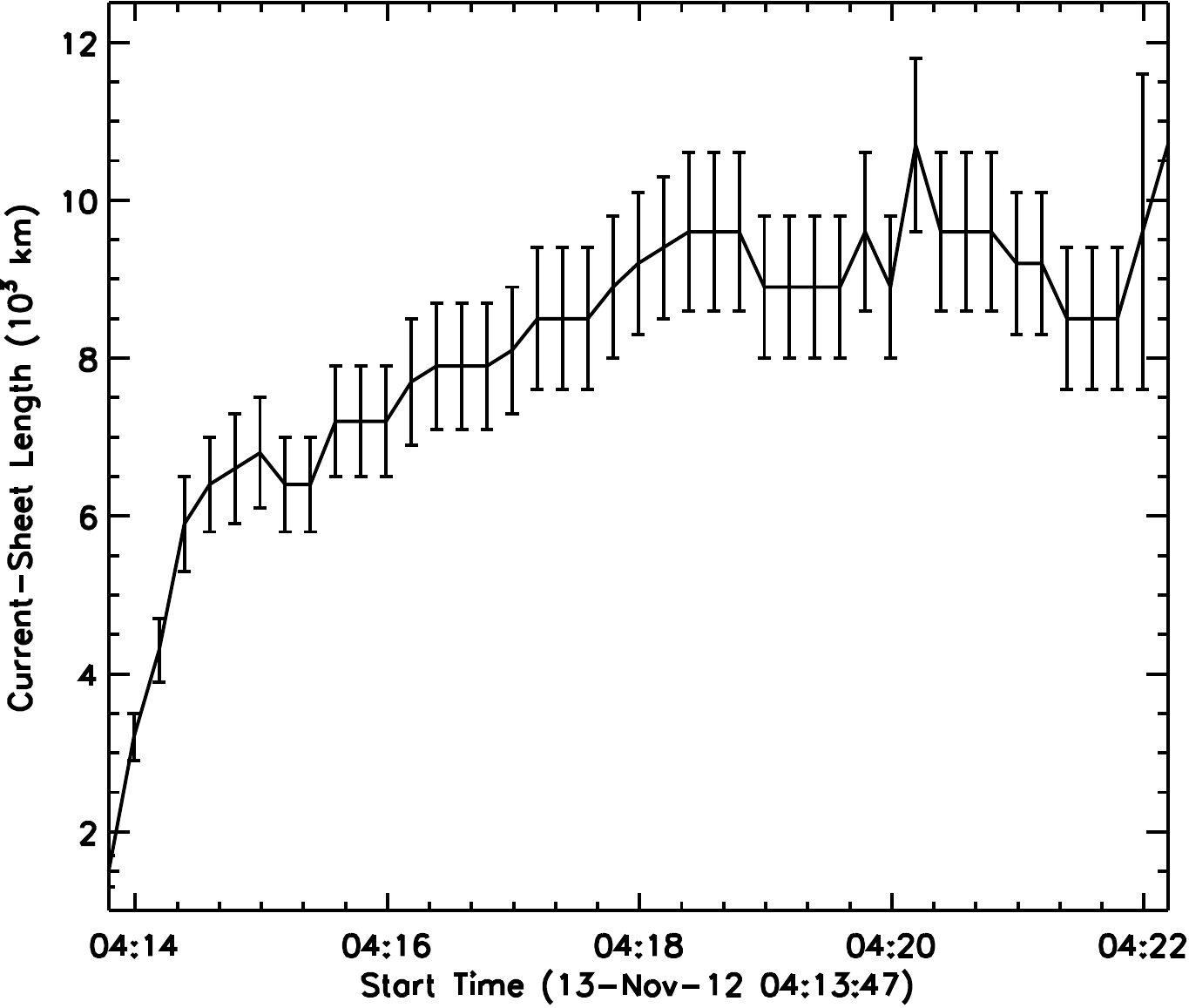}
			\caption{Progression of the measured length of the breakout-reconnection current sheet in the eruption of jet J9, from when the current sheet first became discernible until its disruption.} \label{fig3d}
		\end{figure}

	Figure \ref{fig3d} is the time plot of the length of the bright, flat current-sheet front of the rising minifilament.  This length was measured from each frame of the 12-second-cadence AIA 171 \AA\ movie, starting with the first frame in which the bright front could be discerned (at 04:13:37 UT) and ending with the last frame (at 04:22:11 UT) before the current-sheet flat front began breaking up.  The uncertainty of each measured length was estimated by eye, and is shown as an error bar on each plotted point in Figure \ref{fig3d}.  The plot shows that the current sheet quadrupled in length to about 6,000 km during its first minute of growth, grew to its maximum length of about 9,000 km over the next four minutes, and then fluctuated in length around 9,000 km until it started breaking up three minutes later, at 04:22:23 UT.  Evidently, the breakout-reconnection external current sheet was present from the start of the base-interior brightening and slow rise onward but was too dim and too short to be noticeable and measurable until the minifilament’s slow-rise speed surpassed a few kilometers per second toward the end of the eruption’s slow-rise phase (as the eruption approached, or began entering into, its fast-rise phase).

	%

			\subsection{\textit{Results from Table \ref{tab:list} as a Whole}}\label{all}
		
		From the minifilament slow-rise start and end times in Table \ref{tab:list}, the duration of the slow rise ranged from 4 minutes (J7) to 30 minutes (J9) and averaged 15 minutes for our 10 eruptions.  The median slow-rise duration was 12 minutes.
		 
		 The slow-rise and jet bright point start times in Table \ref{tab:list}  show that the minifilament’s slow rise was observed to start before the jet bright point in each of our 10 jet eruptions, except perhaps in J7.  For J7, the slow-rise start time is 03:27:11 $\pm$ 90 s and the jet bright point start time is 03:28:11 $\pm$ 24 s, giving the uncertainty ranges an overlap of 44 seconds in which the jet bright point is allowed to have started slightly before, simultaneously, or slightly after the minifilament started rising.  From the time-distance plot in Figure \ref{fig2a}  for the J7 minifilament, we judge that it is more likely that the slow rise started during the 90 seconds before the start-time line at 03:27:11 UT than during the 90 seconds after 03:27:11 UT.  Thus, the start times of the slow rise and the jet bright point in Table \ref{tab:list}  indicate that in at least 9 of our 10 jet eruptions, and probably in J7 as well, the runaway internal reconnection that made the jet bright point set in after the eruption of the minifilament had started, i.e., that, at first, the eruptive instability did not involve internal runaway reconnection.  This finding from 10 quiet-region jets observed on the central disk is consistent with the corresponding finding by \cite{moore18} from 15 coronal-hole jets observed near the limb.  In 14 of those 15 jets the jet bright point started after the minifilament started rising, and in the other jet the jet bright point started simultaneously with the rise of the minifilament to within the time resolution of the movies.
		
		     In 8 of our 10 jet eruptions, all except J3 and J4, the start of the base-interior brightening was the first sign that runaway breakout reconnection of the jet-base anemone’s erupting lobe had started, making the base-interior brightening by adding new hot loops to the anemone’s non-erupting lobe.  In these 8 eruptions, the other product of that reconnection, the spire, did not become discernible until after the base-interior brightening could be seen.  In J3 the base-interior brightening and spire became discernible at the same time.  In J4 the spire became discernible about two minutes before the base-interior brightening.  Thus, consistent with the corresponding finding from \cite{moore18}, we find that in quiet-region coronal jets, with few exceptions, the start of the base-interior brightening leads the start of the spire as the first discernible sign that runaway breakout reconnection has started.
		
		     In 8 of the 10 jet eruptions, all except J2 and J9, the minifilament started rising before the base-interior brightening started.  In these 8 eruptions, the base-interior brightening started in the slow rise, except for J1 in which the base-interior brightening started about two minutes into the fast rise.  Also in these 8 eruptions, the jet bright point started before the base-interior brightening in 6 eruptions (J1, J3, J4, J6, J7, J10) but started after the base-interior brightening in 2 eruptions (J5, J8).  In J2 and J9, the base-interior brightening started simultaneously with the start of the minifilament’s slow rise, and the jet bright point started later.  In each of the 10 eruptions except J9, the jet bright point started during the slow rise.  In J9, the jet bright point started simultaneously with the start of the fast rise at the end of the slow rise.  Thus, consistent with \cite{moore18} for coronal-hole jets, our observations indicate that in a majority (8/10) of quiet-region jets, the eruption is initiated by an MHD instability that does not involve runaway reconnection at first but leads to runaway internal reconnection and runaway breakout reconnection that start separately later in the slow-rise phase or early in the fast-rise phase of the eruption and (presumably) help make the eruption more explosive.  Also consistent with \cite{moore18} for coronal-hole jets, our observations indicate that in a minority (2/10) of quiet-region jets, runaway breakout reconnection participates in the eruptive instability throughout the eruption onset, at and after the discernible start of the minifilament’s slow rise.

		     Table \ref{tab:list} shows that in 7 of our 10 jet eruptions (J1, J3, J4, J5, J6, J8, J10), the motion of the minifilament’s slow rise projected on the plane of the sky (its proper motion) had become fast enough to be discerned and measured before any evidence of runaway reconnection became discernible.  In J2 and J9, evidence of runaway reconnection, the base-interior brightening, became discernible simultaneously with the start of the minifilament’s slow rise.  In J7, within the overlap of the uncertainty in the two start times, evidence of runaway internal reconnection, the jet bright point, became discernible perhaps simultaneously with the start of the minifilament’s slow rise.  In all 10 eruptions, the proper-motion speed of the rising minifilament at the end of the slow rise (start of the fast rise) was measured to be faster than at the start of the slow rise.  At the start of its slow rise, the minifilament’s proper-motion speed was about 1 \kms\ in 7 eruptions (J1, J2, J3, J5, J6, J8, J9), about 3 \kms\ in 2 eruptions (J7, J10), and about 8 \kms\ in 1 eruption (J4).  The average minifilament slow-rise start  speed is 2.2 $\pm$ 2.0 \kms, which is generally faster than the horizontal flow speed in granules and supergranules, which have a combined speed typically less than 1 \kms\ \citep{leighton63,simon64,beckers77}. [In only one of our ten jets (J3) is the rising minifilament’s measured proper-motion speed less than 1.0 \kms\ (0.5 $\pm$ 0.3 \kms) at the start of the minifilament's slow rise.  The speed of the horizontal outflow in granules is typically about 0.25 \kms\ \citep{beckers77}.  The speed of the horizontal outflow in supergranules is typically about 0.4 \kms\ \citep{simon64,beckers77}.  So, in the part of a granule in which the granule’s horizontal outflow is in the same direction as the supergranular horizontal flow on which the granule rides, the combined speed is typically about 0.65 \kms.  Because in each of our ten jets, except J3, the minifilament’s measured proper-motion speed (a lower bound on the minifilament’s actual rising speed) was found to be faster than 0.65 \kms\  at the start of the minifilament's slow rise, and because the minifilament continued to rise and erupt, we take the measured initial proper-motion rising speed of the minifilaments in our ten jets, being generally greater than 1.0 \kms, to be an indication that the minifilament-carrying magnetic field had been rendered unstable before the discernible start of the minifilament’s slow rise and its eruption was already underway at the discernible start of the minifilament’s slow rise.] 
		     
		     At the end of the slow rise (start of the fast rise) in these 10 eruptions, the proper-motion speed of the rise was faster than at the start by factors of 2 (J4, J10) to 15 (J8).  So, for each of these 10 eruptions, because no sign of runaway reconnection started before the discernible start of the minifilament’s slow rise, and because the proper-motion speed is a lower bound on the 3D speed of the rising minifilament, the slow-rise, jet bright point, base-interior brightening, and spire start times together with the measured proper-motion speed in Table \ref{tab:list} indicate that when either runaway internal reconnection or runaway external (breakout) reconnection became appreciable, the eruption had already started and had attained an outward bulk speed of at least $\sim$ 1 \kms.
		
	\section{SUMMARY AND DISCUSSION}\label{discussion}

	          The present paper is a follow-up to the paper by \cite{moore18} that similarly examined the onset of the driving magnetic explosion in the magnetic-anemone base of coronal jets.  Because \cite{moore18} studied the eruption in only coronal jets observed near either polar limb, the actual start of the slow rise of the minifilament could have been hidden behind the chromosphere/transition-region spicule forest along the line of sight to the pre-eruption minifilament.  That is, if the top of the pre-eruption minifilament was below the top of the spicule forest, the minifilament would have to have started rising before it could be seen emerging above the spicule forest.   Therefore, in jet eruptions in which the base-interior brightening or the jet bright point was observed to start before the first detection of the rising minifilament, it was uncertain whether the runaway reconnection that made the base-interior brightening or the jet bright point was part of the eruption mechanism at the very start of the minifilament eruption, or, instead, before the base-interior brightening or jet bright point had stared, the minifilament started erupting by an MHD instability not involving runaway reconnection\footnote{Of the 15 near-limb polar jets studied by \cite{moore18}, the observations definitely show that the minifilament was rising before either the base-interior brightening or the jet bright point started in 9 jet eruptions.  Due to possible obscuration of the minifilament by the spicule forest, the observations allow that the minifilament started rising first in the other 6 jet eruptions as well.}.  For this reason, the near-limb observations studied by \cite{moore18} left open the possibility that in quiet regions and coronal holes nearly all jet-producing minifilament eruptions, instead of only a moderately large majority, are started by MHD instability that does not involve runaway reconnection at first but soon leads to both the runaway reconnection that makes the base-interior brightening and the runaway reconnection that makes the jet bright point.   
	
	The present study avoided the near-limb obscuration of the minifilament at eruption onset by studying jet eruptions observed on the central disk, no more than about 30\degree\ from disk center.  We examined the eruptions of the 10 central-disk quiet-region jets that were randomly selected for the studies of \cite{panesar16b,panesar17}.  For each of these eruptions, \cite{panesar16b,panesar17} found from HMI magnetograms that the minifilament magnetic field was built and triggered to erupt by gradual cancelation of opposite-polarity flux driven together at the underlying PIL by photospheric convection.  As in most eruptions of filaments and minifilaments, each of these 10 minifilament eruptions began with a slow rise of uneven, steady or gradually increasing speed and then more rapidly accelerated into the fast-rise phase of the eruption.  For each of these 10 jet eruptions, by stepping through the 12-second-cadence AIA 171 \AA\ movie frame by frame, we visually identified the start and end times of the minifilament’s slow rise and the start times of the jet bright point, base-interior brightening, and spire.  Based on the intensity-time plots and intensity time-distance plots in Figures 3, 6, and 9 respectively for Jets 1, 7, and 9, we believe these times are accurate to within their visually-estimated uncertainties listed in Table \ref{tab:list}.  From the AIA 171 \AA\ movie of each eruption, we made an intensity time-distance plot from which the minifilament’s proper-motion speed was measured at the start of the slow rise and at the end of the slow rise (start of the fast rise).
	
	     \cite{adams14} and \cite{sterling15} and several related papers \citep[e.g.][]{hong11,shen12,huang12,panesar16b,sterling17,mcglasson19}, including the present paper, take the view that minifilament eruptions that make coronal jets are miniature versions of filament eruptions that have breakout reconnection and make CMEs.  In this view, the slow rise and further eruption of the minifilament correspond to those of the filament eruption, the jet bright point is a miniature of the flare arcade that is made by runaway internal reconnection under the erupting filament and straddles the PIL of the pre-eruption filament, and the base-interior brightening is in a miniature of the hot new loops made by runaway external (breakout) reconnection of the erupting filament-enveloping arcade with encountered oppositely directed field.  \cite{moore18} supports this interpretation with evidence that the onsets of minifilament eruptions in polar coronal jets have an observed diversity similar to that laid out by \cite{moore06} for CME-making filament eruptions that have breakout reconnection.  \cite{moore18} present evidence that, as in the onsets of CME-producing filament eruptions that have breakout reconnection, (1) the minifilament often starts rising by MHD instability before the start of either the jet bright point made by runaway internal reconnection or the base-interior brightening made by runaway breakout reconnection, but (2) sometimes the jet bright point and/or the base-interior brightening start nearly simultaneously with the start of the minifilament’s slow rise, indicating that runaway reconnection is part of the mechanism that starts the eruption or sets in very soon after the eruption starts.  Our observations agree with \cite{moore18} in showing to about the same extent that the onsets of nearly all quiet-region jet-producing filament eruptions are in accord with the eruption being a miniature version of a CME-producing filament eruption that has breakout reconnection.  But, more importantly, from our clearer view of the minifilament at eruption onset, we find evidence against a major inference of \cite{moore18}.
	
	\cite{moore18} tacitly assume that, at the interface between two oppositely-directed magnetic fields, an extensive current sheet has to be built up before runaway reconnection can start there.  For their 15 jet eruption onsets, in a  majority (9/15) the rising minifilament was detected before the start of the base-interior brightening, and in the remaining large minority (6/15) the base-interior brightening started either before or simultaneously with the first detection of the rising minifilament.  In a large majority (13/15) the first detection of the jet bright point was last, after the first detections of the rising minifilament and base-interior brightening.  Of the two eruptions in which the jet bright point did not start last, one started simultaneously with the first detection of the rising minifilament and before the base-interior brightening started, and the other started after the first detection of the rising minifilament and simultaneously with the base-interior brightening.  For eruptions in which the base-interior brightening started after the first detection of the rising minifilament, \cite{moore18} reasoned (1) that prior to the start of the minifilament’s slow rise, in the magnetic-null region between the ambient field and the outside of the pre-eruption magnetic lobe enveloping the minifilament, there was either still a null point or the field around the null point was not yet sufficiently compressed to form a current sheet of enough extent for runaway breakout reconnection to have yet started there, and hence (2) that the base-interior brightening-making runaway breakout reconnection started there later when the erupting minifilament-carrying lobe had compressed the null-region field enough to make a current sheet there of enough extent to start runaway breakout reconnection.  For eruptions in which the base-interior brightening started before or simultaneously with the first detection of the rising minifilament, \cite{moore18} reasoned that pre-eruption photospheric-convection-driven evolutionary bulging of the minifilament-holding lobe, slower rising of the lobe than at the start of the eruption’s slow rise, had gradually built an external null-point current sheet that, before the start of the minifilament’s slow rise, became large enough to start runaway breakout reconnection.  \cite{moore18} tacitly also surmised that before the start of the minifilament’s slow rise there could be a gradually-built extensive pre-eruption internal current sheet, between the pre-eruption lobe’s legs under the minifilament, and that in most eruptions this current sheet is not large enough to start runaway reconnection there until it is built large enough by the eruption.
	
	     So, for at least the large minority of jet eruptions in which the base-interior brightening starts simultaneously or nearly simultaneously with the slow rise of the minifilament, \cite{moore18} surmised that photospheric-convection-driven gradual magnetic evolution prior to eruption onset builds (1) an extensive external current sheet in the magnetic null region between the pre-eruption lobe and the ambient field and (2) an extensive internal current sheet between the pre-eruption lobe’s legs under the minifilament.  See the first drawing in Figure 17 of \cite{moore18}.  Our 10 central-disk quiet-region jet eruptions show evidence from which we surmise that the external and internal current sheets large enough to start base-interior-brightening-making and jet-bright-point-making runaway reconnection (1) are much smaller than expected by \cite{moore18} and (2) are seldom, if ever, built even that large by pre-eruption evolution  because they require the faster building by the eruption itself to be built that large and larger.

	     Our evidence and reasoning are the following.  In the AIA 171 \AA\ images of the eruption of jet J9, from late in the slow rise to early in the fast rise, the direction of view of the erupting minifilament happens to clearly show the growth and demise of what is evidently the breakout-reconnection current sheet viewed edge on at the front of the filament (Figure \ref{fig3b} and MOVIE3).  Because a narrow jet spire is seen to extend from the top end of this feature (starting at about 04:15 UT when the minifilament-front feature has more than half its final length), we are confident that this feature is the breakout-reconnection current sheet.  In the AIA 171 \AA\ movie, the current sheet first shows perceptible brightness and length starting at 04:13:47 UT, late in the eruption’s slow rise, which was from 03:50:35 to 04:20:00 UT.  The current sheet rapidly grew to about half its final length in about a minute, and then grew more slowly and fluctuated in length until its disruption at about 2 minutes into the eruption’s fast rise.  After the initial burst of growth, the length of the current sheet is 5-10 x 10$^3$ km, about a quarter of the diameter of the jet-base anemone (Figure \ref{fig3b}).  In proportion to the jet base, this is of the order of the length envisioned by \cite{moore18} for the external current sheet and the internal current sheet that they surmised to have formed gradually before the onset of eruption in a large minority of jets, as is depicted in the first drawing of their Figure 17.  In the AIA 171 \AA\ movie of J9, the breakout current sheet does not become large enough and bright enough to be seen and then grow to a length of this order until late in the slow rise of the eruption, when the 3D speed of the erupting minifilament was probably at least a few kilometers per second (Table \ref{tab:list}).  But, the base-interior brightening in J9 started at the start of the slow rise.  From these observations, we surmise that there was a breakout-reconnection current sheet present and undergoing runaway breakout reconnection from the start of the slow rise (when the 3D speed of the rising minifilament was at least $\sim$1 \kms) and that the current sheet was not made large enough to be discerned and measured in the AIA 171 \AA\ movie until the eruption sped up to a few times faster late in the slow rise.
	
	     In the AIA 171 \AA\ movies of our 10 random central-disk quiet-region jet eruptions, we could not visually detect that the slow rise had started until the proper-motion speed of the minifilament exceeded at least$\sim$ 1 \kms\ (Table \ref{tab:list}).  Also, in none of our 10 eruptions were the signatures of the start of runaway reconnection (the start of the base-interior brightening and the start of the jet bright point) observed before the visually-discerned start of the minifilament’s slow rise.  From these two observations, we surmise that in a large majority of quiet-region jet eruptions the actual start of the eruption’s slow rise is at least shortly before the start of any runaway reconnection, and hence the eruption starts by an MHD instability that does not at first involve runaway reconnection.

	     	\cite{panesar16b,panesar17,panesar18a} and \cite{mcglasson19}  have found that at least a large majority of quiet-region coronal jets are triggered by flux cancelation at the PIL under the minifilament.  From this and the slow-rise speeds in Table \ref{tab:list}, we infer that usually the flux cancelation first renders the pre-eruption lobe unstable to an MHD eruptive instability not involving runaway reconnection, and that when the rise speed exceeds at least $\sim$1 \kms, the eruption begins driving runaway external and internal reconnection fast enough to start making the base-interior brightening and the jet bright point. 
	
	     Because jet-making minifilament eruptions in quiet regions and coronal holes are evidently miniature versions of CME-making filament eruptions that start the same way \citep{sterling18}, and because it appears that at least a large majority of quiet-region jet-making minifilament eruptions start by an MHD instability that results from flux cancelation at the PIL under the minifilament and that does not involve runaway reconnection, we expect that many CME-making filament eruptions start by an MHD instability that results from flux cancelation under the filament and that does not involve runaway reconnection.
	
	     We suppose the observed timing and growth of the base-interior brightening and the external current sheet in J9 to be more or less typical of both the base-interior-brightening-making and the jet-bright-point-making runaway reconnection in most quiet-region coronal jet eruptions.  From that, and from the measured slow-rise speeds in Table \ref{tab:list}, we surmise that in most quiet-region jet eruptions the base-interior brightening-making and jet bright point-making runaway reconnections start at their respective current sheets when the extent of the current sheet is no more than $\sim$ 1\% of the span of the jet-base anemone and the slow-rise speed is at least $\sim$ 1 \kms, and that the reconnection keeps the current sheet from growing to $\sim$ 10\% of the span of the jet-base anemone until the slow-rise speed has further increased by at least a factor of 2 or so.\footnote{The estimates of $\sim$ 1\% and $\sim$10\% come from measurements of AIA 171 \AA\ images of jet J9.  From these images (from which we obtained the intensity time-distance map in Figure \ref{fig3c}d for the breakout current sheet), we found that the measured length of the breakout current sheet (Figure \ref{fig3d}) was $\sim$ 1\% of the major diameter of the jet-base anemone (the full extent of which is illuminated in Figure \ref{fig3b}d) when the current sheet first became discernible, and was of the general order of magnitude of 10\% (it was about 25\% for J9) at the end of the current sheet’s growth.}
	
		\begin{figure}
		\centering
		\includegraphics[width=\linewidth]{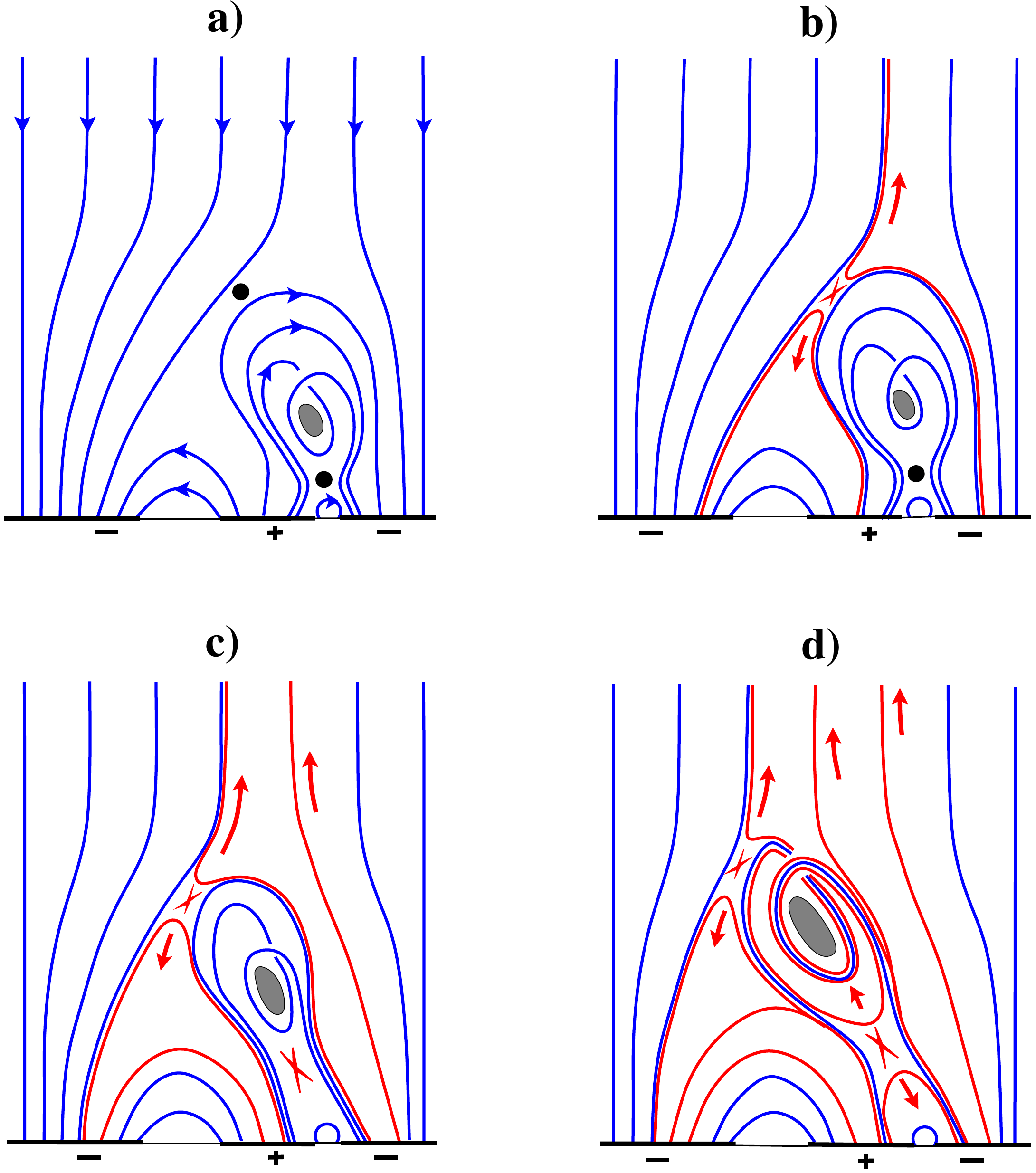} 	
		\caption{Schematic depiction of the driving magnetic explosion in quiet-region and coronal-hole coronal jets in which the base-interior brightening starts before the jet bright point (revised from \citealt{moore18}).  Thick horizontal black lines are patches of magnetic flux; the polarity of each patch is given by a plus or minus sign.  Blue curves are field lines that have not undergone reconnection.  Red curves are field lines that have undergone reconnection.  Black dots between opposite-direction blue field lines are magnetic null points or current sheets of no appreciable extent.  Red crosses mark sites of ongoing runaway reconnection.  Red arrows depict plasma outflow along reconnected field lines.  The gray blob wrapped in a coiled field in the core of the explosive lobe is the cool-plasma filament viewed end-on.  In panel (a), arrow heads show the direction of the field lines, the explosive lobe has not yet started erupting, and no runaway reconnection has yet started at either null-point current sheet.  The rise of the minifilament starts and leads to the start of runaway breakout reconnection at the external current sheet in the time interval between panels (a) and (b), and both of these actions are continuing in panels (b)-(d).  Runaway tether-cutting reconnection at the internal current sheet is starting in panel (c) and is continuing in panel (d), unleashing the minifilament eruption to increasingly more strongly drive the spire-producing breakout reconnection. } \label{fig4}
	\end{figure} 

	Figure \ref{fig4} corresponds to Figure 17 of \cite{moore18}.  Their drawings are for the pre-eruption magnetic field and its explosion in the large minority of their near-limb jet eruptions in which the base-interior-brightening-making runaway breakout reconnection starts at or near the start of the minifilament’s slow rise and before the start of the jet-bright-point-making runaway internal reconnection.  Likewise, the drawings in Figure \ref{fig4}  are for jet eruptions in which the base-interior brightening starts before the jet bright point, but our panel-(a) drawing depicts the pre-eruption set up that we expect for at least a large majority of quiet-region and coronal-hole jets whether or not the base-interior brightening-making reconnection starts before the jet bright point-making reconnection.  The drawings are the same as those in Figure 17 of \cite{moore18} except for the lengths of the pre-runaway-reconnection external and internal current sheets.  \cite{moore18} assumes that, before runaway reconnection can start at either current sheet, the current sheet has to have a length of $\sim$ 10\% of the span of the jet-base anemone.  In contrast, from the onsets of our 10 central-disk jet eruptions, especially the eruption of J9, we surmise that in most jet eruptions, before and at eruption onset these current sheets are much smaller than that, the length of each being no more than $\sim$ 1 \% of the span of the jet base.  So, in Figure \ref{fig4}, each of the two current sheets, before the onset of runaway reconnection there, is represented by a dot instead of by the thick line segment in Figure 17 of \cite{moore18}.
	
	Few, if any, coronal jets in quiet regions and coronal holes are observed to be made in the way that was expected by \cite{yokoyama95}, by the anemone-building reconnection of the outside of an emerging bipole with ambient far-reaching magnetic field.  Instead, it is observed that nearly all are made by a minifilament eruption that drives the reconnection of its enveloping lobe of the jet base with the ambient far-reaching field \citep{sterling15,panesar16b,panesar18a,mcglasson19}.  Our observations offer an explanation for the absence of jets directly made by external reconnection of an emerging bipole.  In Table \ref{tab:list}, the slow-rise start and end times and start and end speeds and the spire start times together imply that for the breakout reconnection to be vigorous enough to make a detectable spire, the rising of the minifilament eruption that drives the reconnection has to be faster than $\sim$ 1 \kms.  This suggests that the basic reason that an emerging bipole seldom or never makes a detectable jet by reconnection with ambient far-reaching field is that emerging bipoles seldom or never emerge fast enough to do that.
	
	     Finally, from our findings for the onset of runaway reconnection and current-sheet size in jet-producing minifilament eruptions, we expect the following for current-sheet building in any astrophysical setting of magnetic field in low-beta plasma.  We expect that: (1) no current sheet of appreciable extent compared to the extent of the overall magnetic field arrangement can be built at the separatrix or quasi-separatrix between two domains of magnetic field of different direction by quasi-static gradual convergence of the two domains, (2) a current sheet of appreciable extent can be built there only dynamically by an MHD convulsion of the field, such as in the minifilament eruptions that build sizable current sheets in the production of coronal jets, and (3) for quasi-static slow converging of the opposite domains, the current sheet does not attain an appreciable extent because the reconnection tears down the current sheet as fast as the converging builds it up.  Therefore, in CME-making filament eruptions that have breakout reconnection, in addition to expecting that the eruption starts by an MHD instability without any runaway reconnection at first, we also expect that no sizable external or internal current sheet is present at or before the start of the filament’s slow rise.
	
It is worth mentioning that because all of our  quiet region jets are from 2012,  we do not know if the initiation of jet eruption mechanism changes over different phases of a solar cycle.  It will be interesting to investigate a large sample of jets observed from different phases of a solar cycle.	
	
	\acknowledgments
	NKP acknowledges support from NASA’s \sdo/AIA contract (NNG04EA00C) to LMSAL. AIA is an instrument onboard the Solar Dynamics Observatory, a mission for NASA’s Living With a Star program. A.C.S and R.L.M were supported by funding from the Heliophysics Division of NASA's Science Mission Directorate through the heliophysics Guest Investigators Program, and by the \Hinode\ Project. We are indebted to the \sdo/AIA and \sdo/HMI teams for providing the high resolution data. \sdo\ data are courtesy of the NASA/\sdo\ AIA and HMI science teams.

\bibliographystyle{aasjournal}

\end{document}